\documentclass[twocolumn]{aastex61}

\usepackage{rotating}

\shorttitle{Complete Survey of $z \sim 5.5$ Quasars.}
\shortauthors{Yang et al.}
\begin{document}

\title{Filling in the Quasar Redshift Gap at $z \sim 5.5$ II: A Complete Survey of Luminous Quasars in the Post-Reionization Universe}

\correspondingauthor{Jinyi Yang}
\email{jinyiyang@as.arizona.edu}

\author{Jinyi Yang}
\affil{Steward Observatory, University of Arizona, 933 N Cherry Ave, Tucson, AZ, USA}
\affil{Kavli Institute for Astronomy and Astrophysics, Peking University, Beijing 100871, China}

\author{Feige Wang}
\affil{Department of Physics, University of California, Santa Barbara, CA 93106-9530, USA}
\affil{Kavli Institute for Astronomy and Astrophysics, Peking University, Beijing 100871, China}

\author{Xiaohui Fan}
\affil{Steward Observatory, University of Arizona, 933 N Cherry Ave, Tucson, AZ, USA}

\author{Xue-Bing Wu}
\affil{Kavli Institute for Astronomy and Astrophysics, Peking University, Beijing 100871, China}
\affil{Department of Astronomy, School of Physics, Peking University, Beijing 100871, China}

\author{Fuyan Bian}
\affil{European Southern Observatory, Alonso de C\'ordova 3107, Casilla 19001, Vitacura, Santiago 19, Chile}

\author{Eduardo Ba\~nados}
\affil{The Observatories of the Carnegie Institution for Science, 813 Santa Barbara Street, Pasadena, California 91101,USA}

\author{Minghao Yue}
\affil{Steward Observatory, University of Arizona, 933 N Cherry Ave, Tucson, AZ, USA}

\author{Jan-Torge Schindler}
\affil{Steward Observatory, University of Arizona, 933 N Cherry Ave, Tucson, AZ, USA}

\author{Qian Yang}
\affil{Department of Astronomy, School of Physics, Peking University, Beijing 100871, China}
\affil{Kavli Institute for Astronomy and Astrophysics, Peking University, Beijing 100871, China}

\author{Linhua Jiang}
\affil{Kavli Institute for Astronomy and Astrophysics, Peking University, Beijing 100871, China}

\author{Ian D. McGreer}
\affil{Steward Observatory, University of Arizona, 933 N Cherry Ave, Tucson, AZ, USA}

\author{Richard Green}
\affil{Steward Observatory, University of Arizona, 933 N Cherry Ave, Tucson, AZ, USA}

\author{Simon Dye}
\affil{School of Physics and Astronomy, University of Nottingham, University Park, Nottingham, NG7 2RD, UK}



\begin{abstract}
We present the final results from our survey of luminous $z \sim $ 5.5 quasars. This is the first systematic quasar survey focusing on quasars at $z \sim$ 5.5, during the post-reionization epoch. It has been challenging to select quasars at $5.3 < z < 5.7$ using conventional color selections, due to their similar optical colors to those of late-type stars, especially M dwarfs. We developed a new selection technique for $z \sim$ 5.5 quasars based on optical, near-IR, and mid-IR photometry, using data from the Sloan Digital Sky Survey (SDSS), PanSTARR1 (PS1), the UKIRT Infrared Deep Sky Surveys -- Large Area Survey, the UKIRT Hemisphere Survey, the VISTA Hemisphere Survey, and the {\it Wide Field Infrared Survey Explorer} ({\it WISE}), covering $\sim$ 11000 deg$^2$ of high galactic latitude sky. In this paper, we present the discovery of 15 new quasars at $z\sim 5.5$. Together with results from \cite{yang17}, our survey provides a complete, flux-limited sample of 31 quasars at $5.3 \le z \le 5.7$. 
We measure the quasar spatial density at $z \sim 5.5$ and $M_{1450} < -26.2$. Our result is consistent with the rapid decline of quasar spatial density from $z = 5$ to 6, with $k=-0.66 \pm 0.05$ ($\rho(z) \propto 10^{kz}$). 
In addition, we present a new survey using optical colors only from the full PS1 area for luminous quasars at $z = 5.0 - 5.5$, beyond the SDSS footprint, and report the preliminary results from this survey, including 51 new quasars discovered at $4.61\le z \le5.71$.

\end{abstract}




\section{Introduction} 
At high redshift (z $>$ 5), quasars, the most luminous non-transient light sources, are important tracers for the study of the early Universe.  
However, their discoveries have been challenging,  due to a combination of  low spatial density and high contaminants from cool dwarfs when using color selection. 
More than 300,000 quasars have been discovered \citep[e.g.][]{schneider10,paris14,paris16}, while only $\sim$ 290 quasars are known at $z >$ 5 \citep[e.g.][]{fan01a,fan06,willott10b,mortlock11,venemans13,venemans15,kashikawa15,wu15,matsuoka16,wang16,jiang16,wang17,mazzucchelli17,yang17,jeon17,banados18,matsuoka18a,matsuoka18b}. 
In particular, in the distribution of quasar redshift, there is a striking gap at $z \sim 5.5$
in the redshift distribution of known quasars.
This is caused by the fact that quasars at $z \sim 5.5$ have similar optical colors to those of late-type stars (see the redshift distribution in Section 4, and  Figure 2) when using
optical broad-band colors alone \citep[][hereafter Paper I]{yang17}. 
Until recently, only $\sim$ 30 known quasars have been discovered in this redshift gap ($5.3 \le z \le 5.7$) \citep[e.g.][]{stern00, cool06, romani04, douglas07, matute13, banados16}. These quasars were selected from a number of different methods and over a very wide magnitude range (17.5 $< z <$ 26 mag) with complex selection effects, 
thus can not be used as a complete sample for statistical studies. This has been a main limitation in understanding the evolution of quasar population and the properties of the intergalactic medium (IGM) at $5<z<6$.

Using absorption spectra of high redshift quasars, observations of the Gunn-Peterson effect have suggested that reionization is largely completed by $z\sim$ 6, possibly with a tail to $z\sim$ 5.5 \citep{becker15, fan06, mcgreer15}.  
The physical conditions of the post-reionization IGM at $z = 5 - 6$ provide the basic boundary conditions of reionization models, such as the IGM temperature, photon mean free path, metallicity and the impact of helium reionization \citep[e.g.][]{bolton12}. They place strong constraints on the reionization topology as well as on the chemical feedback by early galaxies and the sources of reionization. 

This redshift range is also a key epoch to the study of evolution of quasar spatial density and black hole (BH) growth. The quasar spatial density or quasar luminosity function (QLF) and BH mass have been measured at $z \sim$ 5 and 6 \citep{mcgreer13, jiang08,willott10a,willott10b,kashikawa15,yang16}. \cite{mcgreer13} measured the quasar spatial density at $z \sim$ 4, 4.9 and 6, and found that luminous quasars decline more rapidly, roughly at twice the rate, at $z \sim 5 $ to 6 than the decline from $z \sim 4$ to 5 (see also \cite{jiang16}).
However, the exact evolution of quasar spatial density from $z$ = 5 to 6 has not been confirmed because of the small size and high incompleteness of the existing quasars between those two redshift bins at which well-established studies exist. 
In addition, \cite{willott10a} suggested a rapid BH mass growth phase after $z \sim$ 6. \cite{trakhtenbrot11} studied the BH growth at $z \sim$ 4.8 and suggest the notion of fast SMBH growth at $z \sim$ 4.8, corresponding to probably the first such phase for most SMBHs. Therefore, a SMBH sample at $z\sim 5.5$ will play a significant role in understanding early SMBH growth. 

To answer the questions posted above, a large, uniformly selected quasar sample at $z \sim 5.5$ is required. This has proven to be challenging. 
Typical $z \gtrsim 5$ quasar selection method is based on the broad optical bands, such as $r,i,z$ bands.  
This method is not effective for quasars at $z \gtrsim 5.3$, when the quasar color track starts to enter the M dwarf locus in the $riz$ color space. 
As shown in Section 2,  $z\sim 5.5$ quasars and much more numerous M dwarfs have almost identical broad-band colors, if only small number of passbands are used. 
Therefore, to avoid the lager number of M dwarf contaminations, previous quasar surveys always used selection criteria that excluded the M dwarf locus in the $riz$ color-color diagram. 
As a result,  $z \sim 5.5$ quasars were rejected at the same time. Known quasar population is highly incomplete at this redshift. 

In Paper I, we presented a new method by combining optical $r-i/i-z$ color-color selection with near-infrared (NIR) and mid-infrared (MIR) colors to select quasars at $z\sim 5.5$.
We carried out a systematic $z \sim 5.5$ quasar survey in a $\sim $ 4500 $deg^{2}$ area, using photometry from the Sloan Digital Sky Survey (SDSS), the UKIRT Infrared Deep Sky Surveys--Large Area Survey \citep[ULAS][]{lawrence07}, the VITSA Hemisphere Survey \citep[VHS][]{mcmahon13} and the {\it Wide Field Infrared Survey Explorer} \citep[{\it WISE},][]{wright10}. Paper I presented the discovery of 17 new $z \sim 5.5$ quasars from this survey. One quasar that was selected using the UKIDSS Hemisphere Survey \citep[UHS,][]{dye18} instead of ULAS was also reported in Paper I as a test of UHS selection, which suggested the feasibility of expanding the survey area out of ULAS.

We have now expanded the survey to a larger area covered by UHS, and completed quasar identification of the SDSS--ULAS/UHS/VHS--{\it WISE} selection over a total of $\sim$ 11000 deg$^2$.  
In this paper, we present the discoveries of 13 new quasars at $z\sim 5.5$ in this new area.
These quasars, combined with objects presented in Paper I, form a complete sample of luminous quasars at $5.3 < z < 5.7$, including 31 objects. Furthermore, we extended the selection to areas outside the SDSS footprint by using the Pan-STARRS1 \citep[PS1;][]{chambers16} dataset, which, at the same time, reaches fainter flux level compared to the sample using SDSS data. 
The selection method and candidate sample are described in Section 2. The observations and result of the survey are presented in Section 3. Using our sample, we calculate the number density of luminous $z\sim 5.5$ quasars in Section 4. In Section 5, we introduce the new survey of $z \sim 5$ and 5.5 quasar using PS1--$J$--{\it WISE} color selection. A summary of our quasar surveys is given in Section 6.
In this paper, we adopt a $\Lambda$CDM cosmology with parameters $\Omega_{\Lambda}$ = 0.7, $\Omega_{m}$ = 0.3, , and H$_{0}$ = 70 $km s^{-1} Mpc^{-1}$. Photometric data from SDSS are reported in the SDSS photometric system \citep{Lupton99}, which is almost identical to the AB system at bright magnitudes; photometric data from PS1 survey are in AB system; photometric data from IR surveys are in the Vega system. All photometric data shown in this paper are corrected for Galactic extinction \citep{schlegel98,schlafly11}.

\section{Selection Method and Photometric Dataset}

In this section, we introduce the selection method and photometric data used for our quasar survey. We  first briefly review the color selection used in the SDSS--ULAS/VHS--{\it WISE} area, the main sample in Paper I (Section 2.1), then discuss the selection extended to the UHS area (Section 2.2). The PS1 $z$ and $y$ data are added into the SDSS--ULAS/UHS/VHS--{\it WISE} color selection for further improvement of efficiency, which is described in Section 2.3. After applying all selection criteria, the final quasar candidate sample is presented in Section 2.5. All photometric data used in our survey is listed in Table 1.

\begin{table}
\caption{Photometric data used in our $z \sim 5.5$ quasar survey.}%T-1
\centering
\begin{tabular}{l l l l}
%\width{0pt}
\hline\hline
  \multicolumn{1}{c}{Survey} &
  \multicolumn{1}{c}{Area ($deg^{2}$)} &
  \multicolumn{1}{c}{Data used} &
  \multicolumn{1}{c}{Depth (5$\sigma$)} \\
\hline
  SDSS & 14000 & u, g, r, i, z & 22.3, 23.3, 23.2, 22.3, 20.8\\
  PS1 & 31012 & z, y & 22.3, 21.4\\
  ULAS & 3700 & J, H, K & 19.6, 18.8, 18.2 (Vega)\\
  UHS & 12700 & J & 19.6 (Vega)\\
  VHS & 20000 & J, H, Ks & 20.2, 19.2, 18.2 (Vega)\\
  WISE & all-aky & W1, W2 & 17.1, 15.7 (Vega)\\
\hline
\end{tabular}
\end{table}

\subsection{Selection Method and Quasar Survey in \cite{yang17}}

The $r-i/i-z$ color - color diagram is a widely adopted selection method previously used for $z \sim 5$ quasar surveys \citep{fan99, richards02, mcgreer13}, as at $z \sim 5$ the Ly$\alpha$ emission line moves into the $i-$band and Lyman series absorption systems begin to dominate the $r-$band. Towards higher redshift ($z > 5.1$), the $i - z$ color becomes redder and enters the M dwarf locus in the $r-i/i-z$ color-color diagram (e.g. Figure 1 in Paper I, also see Figure 2). The continua of M dwarfs later than M4 type  have their SED peak at $\sim$ 8000 -- 12000 $\rm \AA$ \citep[][also Paper I]{kirkpatrick93, mclean03}, seriously contaminating quasar colors in broad band photometry. The locus of M4 -- M8 type dwarfs almost overlaps with the whole region of quasars at $5.3 \le z \le 5.7$ in optical color space. Therefore, it is challenging to select quasars at $z \gtrsim 5.2$ using only optical colors, until at $z \gtrsim 5.7$, at which  the quasar color track moves out of the M dwarf locus. This results in an obvious gap $5.3 \le z \le 5.7$ in the redshift distribution of know quasars (see Figure 4). 

To search for quasars in this redshift distribution gap, we developed a new method by adding NIR and MIR colors (Paper I). 
We started with the $r-i/i-z$ method to restrict quasar redshift within the range of $5.3 \le z \le 5.7$ and then applied the $J - $ W1/ W1 - W2 and $H - K$/$K -$W2 color-color cuts, as shown in Figure 1. 
Based on these colors, we carried out the first systematic $z \sim 5.5$ quasar survey in the $\sim$ 4500 deg$^2$ of SDSS--ULAS/VHS--{\it WISE} area, using SDSS DR10, ULAS DR10, VHS DR3 and ALLWISE data\footnote{http://wise2.ipac.caltech.edu/docs/release/allwise/}. We selected only SDSS point sources (type = 6). We limited our survey area to Galactic latitude $|$b$|$ $>$ 20$^{\circ}$. A $J - K$ color cut was added for further rejection of stars. All selection criteria used for this survey are listed as follows. More details can be found in Paper I. 
\begin{figure*}%f1
\centering
\epsscale{1.2}
\plotone{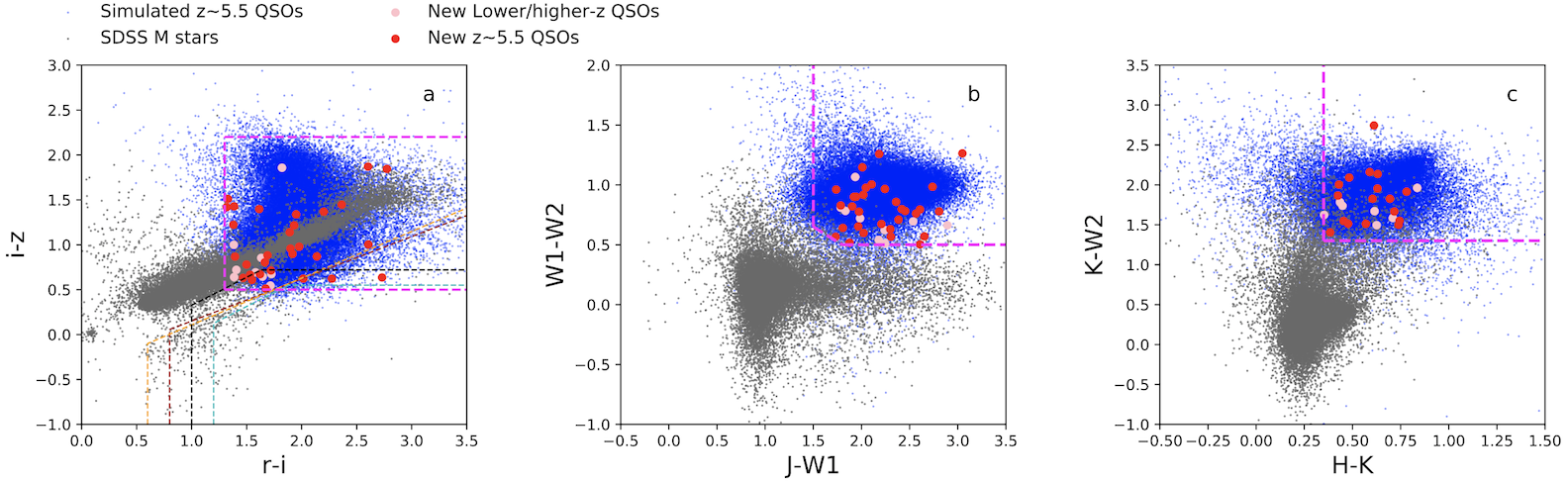}
\caption{Our color-color selection for 5.3 $\le z \le$ 5.7 quasars based on the photometric data in $r,i,z$,$J,H,K$ W1 and W2 bands from SDSS, ULAS and ALLWISE, as described in Paper I. Blue dots denote simulated quasars at 5.3 $\le z \le$5.7 with SDSS z band brighter than 20.8, which is the 5$\sigma$ magnitude limit of SDSS $z$ band. The simulated quasar sample used here is the same sample described in Section 4.2. Grey dots show the locus of SDSS DR10 spectroscopically identified M dwarfs. The purple dashed lines represent our selection criteria, compared with previous $r-i/i-z$ selection criteria for $z \sim$ 5 quasar \citep[][in brown, orange, cyan and black dashed line, respectively]{fan99, richards02, mcgreer13, wang16}. Our method modifies the traditional $r-i/i-z$ color cuts for quasars at $z \sim$ 5 to cover quasars at $z \sim$ 5.5 (selection {\bf a}) and adds the $J-$W1/W1$-$W2 (selection {\bf b}), $H-K$/$K-$W2 (selection {\bf c}) color-color criteria. All new quasars from our survey, in both Paper I and this paper, are represented as red ($z \sim 5.5$ quasars) and pink (quasars at lower/higher redshift, but in the range of $4.5 < z < 5.8$) filled circles. The survey in the UHS area used only selections {\bf a} and {\bf b}, as there is no $H$ and $K$ bands data in UHS.} 
\end{figure*}

% SDSS cuts
\begin{equation}
u>22.3, g > 23.3, z<20.5
\end{equation}
\begin{equation}
S/N(i) >3, S/N(z) >3
\end{equation}
\begin{equation}
g>24.0~ or ~ g-r>1.8
\end{equation}
\begin{equation}
r-i > 1.3
\end{equation}
\begin{equation}
0.5 < i-z < 2.2
\end{equation}

%ULAS/VHS-WISE
\begin{equation}
S/N(W1) \ge 5, S/N(W2) \ge 3
\end{equation}
\begin{equation}
J-W1 > 1.5
\end{equation}
\begin{equation}
W1-W2 > 0.5
\end{equation}
\begin{equation}
W1-W2 > -0.5 \times(J - W1) + 1.4
\end{equation}
\begin{equation}
J - K > 0.8
\end{equation}
\begin{equation}
H - K > 0.35 ~ and ~ K-W2 > 1.3
\end{equation}

After applying these selection criteria, we selected 1339 unique objects, including 1088 objects from ULAS and 284 from VHS (33 overlaps).
Among this candidate sample, there are three known SDSS quasars at $1.96 < z < 5.16$, three known stars and one known faint galaxy. In Paper I, we reported our observations of 93 candidates. Among them, 24 new quasars have been discovered, including 17 quasars at $5.3 \le  z  \le 5.7$.

\subsection{Survey in UHS area}
The SDSS--ULAS/VHS--WISE area is only $\sim$ 4500 deg$^2$. To expand our quasar sample, we need larger area optical/NIR photometric data. The UHS survey 
maps $\sim$ 12,700 deg$^2$ new area in the Northern sky at $0^{\circ} \le $ decl. $\le 60^{\circ}$ beyond ULAS. The UHS survey has a 5$\sigma$ depth of $J$ = 19.6 mag in Vega. The UHS DR1 data has become public in August 2018. We used the preliminary version of UHS data for our candidate selection in September 2016. 
The SDSS-UHS photometric dataset provided $\sim$ 7000 deg$^2$ additional area to our quasar survey. 

For the UHS area survey, we also started with our optically selected candidate sample. We cross-match (2") this candidate sample with UHS and ALLWISE catalogs. Since there is only $J$ band data in UHS, we removed the selection criteria that involve $H $ and $K$ bands, which are equ. (10) \& (11) (i.e. selection {\bf c} in Figure 1). Here we limit candidates to $z <$ 20 mag in order to obtain a relative small candidate sample, which resulted in a sample of 1910 objects that met all selection criteria. Among them, there are eight previously known quasars at $5.07 < z < 5.81$ and one new quasar overlapped with ULAS/VHS sample. 
As pilot observation of the UHS selection, in Paper I, we observed a few candidates and discovered one new quasar J1016+2541 at $z = 5.64$.

\subsection{Adding Pan-STARR1 $z$ and $y$ Data}

Both the ULAS/VHS and UHS candidate samples contain more than 1000 candidates, which are still too many to be spectroscopically identified. We need further selection to reduce the size of the candidate sample. PS1 maps the entire sky at decl. $> -30^{\circ}$ in $g, r, i, z$ and $y$ bands \citep{chambers16}. It has a depth (5 $\sigma$) of 23.2, 23.2, 23.1, 22.3 and 21.2 mag in $g, r, i, z$ and $y$ respectively. The sky coverage and depth make PS1 data an excellent choice for high redshift quasar selection. 
In this section, we only use the $z$ and $y$ band colors to further reject M dwarfs from our SDSS--ULAS/VHS/UHS--{\it WISE} candidate samples.
In principle, the PS1 dataset can also be used for the $riz$ cuts independently. In Section 5, we will introduce the new survey using optical data from PS1 only instead of SDSS.

The narrower PS1 $z$ and $y$ bands centered at 8679 \AA\ and 9633 \AA\ are more sensitive to the difference between spectra of $z \sim 5.5 $ quasars and M4 --  M8 dwarfs than when only using SDSS $i$ and $z$ band. As shown in Figure 2, the $z - y$/$y - J$ color -- color diagram is effective to reject M dwarfs. The M1 -- M3 dwarfs can be rejected by the SDSS $r-i/i-z$ cut already, and the M4 -- M8 dwarfs, which fully overlap with the quasar color track in the $r-i/i-z$ color space, are mostly separated from quasars in the $z-y$/$y-J$ color space. We built the selection criteria based on simulated $z \sim 5.5$ quasars, SDSS DR10 M dwarf sample, the new $z \sim 5.5$ quasar sample and the new M dwarfs from our observations in Paper I. 17 of 18 (94\%) our new $z \sim 5.5$ quasars can be selected by $z-y$/$y-J$ selection \footnote{Quasar J1006-0310 from Paper I at $z = 5.5$ is not included}, while only 6 of 71 (8\%) M dwarfs from Paper I are included. A detailed estimate of the success rate and the completeness are presented in Section 3 and Section 4.2. The selection criteria are listed as following. PS1 $z$ and $y$ magnitudes are Galactic extinction corrected \citep{schlafly11}.

\begin{figure}%f2
\centering
\epsscale{1.1}
\plotone{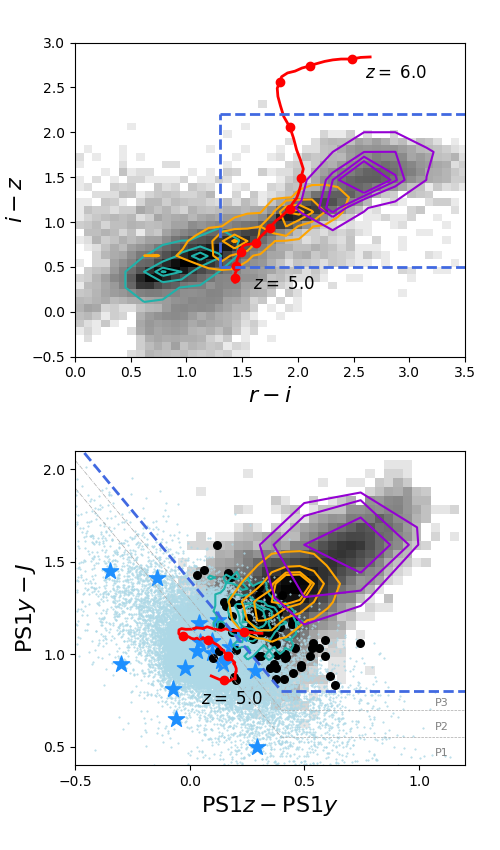} 
\caption{The $r-i$/$i-z$ (SDSS) color-color diagram ({\it Top}) and PS1$z$ $-$ PS1$y$/PS1$y-J$ color-color diagram ({\it Bottom}). The red solid line with red dots are color tracks generated by calculating mean colors of simulated quasars from $z = 5 $ to $z = 6$. Lightblue dots are simulated quasars at $5.3 \le z \le 5.7$. Green, orange and purple contours show the loci of M1--M3, M4--M6 and M7--M9 type M dwarfs. The blue dashed lines represent our selection criteria. In the $r-i$/$i-z$ diagram, the color track has a step of $\Delta z$ = 0.1. Our $riz$ selection restricts quasars to be at $z \sim 5.5$ but also include lots of M dwarfs (mainly M4 and later type) at the same time. On the other hand, the $riz$ cut is useful to exclude early type dwarfs. In the PS1$z$ $-$ PS1$y$/PS1$y-J$ diagram, the red points represent quasars at $z$ = 5.0, 5.3, 5.5, 5.7 and 6. Contours in this diagram are M dwarfs that satisfy the $riz$ color-color cut. Blue stars and black filled circles are $z \sim 5.5$ quasars and M dwarfs from Paper I. Most $riz$ selected M dwarfs are rejected by the $zyJ$ selection. Two dash-dot lines divide our candidate sample into three subsamples with different priority for spectroscopic observations (see details in Section 3).} 
\end{figure}

%Pan-STARR colors for high priority sample.
\begin{equation}
S/N({\rm PS1}z) >3, S/N({\rm PS1}y) >3, S/N(J) >3
\end{equation}
\begin{equation}
 \begin{array}{l}
{\rm PS1}y-J < 0.8 \\
~or~\\
{\rm PS1}y-J < -1.5\times({\rm PS1}z - {\rm PS1}y) + 1.4
 \end{array}
\end{equation}

\subsection{Candidate Sample}
We matched (2") our candidate sample from SDSS--ULAS/VHS/UHS--{\it WISE} selection with PS1 DR1 photometric data and used the photometric data from primary detection (primarydetection = 1). After applying the $z-y$/$y-J$ cuts, we obtained a sample of 208 candidates from ULAS/VHS based and UHS based sample. We visually checked optical images and removed targets with suspicious detections, such as multiple peaked objects or being affected by nearby bright stars. 
The final candidate sample included 54 candidates from the ULAS selection, 10 from the VHS selection and 105 from the UHS selection. There is one overlap between ULAS and VHS selections, confirmed to be a M dwarf. There is one overlap between ULAS and UHS samples, which is quasar J1133+1603 ($z = 5.61$) published in Paper I. Therefore, we finally obtained 167 unique candidates. Among them, there are seven previously known quasars (only one at $z \sim 5.5$), one known galaxy and 28 objects observed in Paper I (22 new quasars and 6 stars).  

\section{Observations and Survey Results}
\subsection{Spectroscopic Identifications}

\begin{sidewaystable}
\centering
\caption{Spectroscopic information of 15 newly identified quasars.} 
\scriptsize
\begin{tabular}{c c c c c c c c c c c c c c c c}
\hline
\multicolumn{1}{c}{Name} &
\multicolumn{1}{c}{Redshift} &
%\multicolumn{1}{c}{SDSS\_$u$} &
%\multicolumn{1}{c}{SDSS\_$g$} &
\multicolumn{1}{c}{SDSS\_$r$} &
\multicolumn{1}{c}{SDSS\_$i$} &
\multicolumn{1}{c}{SDSS\_$z$} &
\multicolumn{1}{c}{$J$} &
\multicolumn{1}{c}{$W1$} &
\multicolumn{1}{c}{$W2$} &
\multicolumn{1}{c}{PS1\_$z$} &
\multicolumn{1}{c}{PS1\_$y$} &
\multicolumn{1}{c}{$M_{1450}$} &
\multicolumn{1}{c}{Instrument} &
\multicolumn{1}{c}{Exp(s)} &
\multicolumn{1}{c}{ObsDate} \\
\hline\hline
SDSS J001232.88+363216.10 & 5.44 & 22.20$\pm$0.12 & 19.92$\pm$0.03 & 19.30$\pm$0.05 & 18.06$\pm$0.06 & 15.76$\pm$0.04 & 15.13$\pm$0.07 & 19.19$\pm$0.01 & 19.19$\pm$0.02 & $-$27.25 & P200/DBSP & 4200 & 2016-09-05\&11\\
SDSS J003414.35+375954.00 & 5.63 & 24.62$\pm$0.53 & 21.84$\pm$0.13 & 19.99$\pm$0.09 & 18.78$\pm$0.11 & 16.73$\pm$0.09 & 15.76$\pm$0.13 & 19.98$\pm$0.02 & 20.11$\pm$0.06 & $-$26.44 & P200/DBSP & 3103 & 2016-09-05\&08\\
SDSS J005656.04+224112.16 & 5.49 & 22.92$\pm$0.36 & 21.03$\pm$0.09 & 19.89$\pm$0.10 & 18.84$\pm$0.10 & 16.92$\pm$0.10 & 16.02$\pm$0.16 & 20.17$\pm$0.03 & 20.10$\pm$0.07 & $-$26.77 & P200/DBSP & 4200 & 2016-09-06\&11\\
SDSS J011455.11+323816.33 & 5.22 & 22.32$\pm$0.18 & 20.94$\pm$0.08 & 19.94$\pm$0.12 & 19.05$\pm$0.13 & 16.80$\pm$0.08 & 16.29$\pm$0.19 & 20.18$\pm$0.03 & 20.01$\pm$0.06 & $-$26.24 & P200/DBSP & 2100 & 2016-09-07\\
SDSS J012053.93+214706.20 & 5.42 & 22.23$\pm$0.17 & 20.73$\pm$0.06 & 19.95$\pm$0.12 & 19.08$\pm$0.12 & 16.63$\pm$0.08 & 15.85$\pm$0.17 & 20.13$\pm$0.03 & 19.97$\pm$0.06 & $-$26.49 & P200/DBSP & 1800 & 2016-09-06\\
SDSS J015745.45+300110.68 & 5.63 & 23.64$\pm$0.64 & 21.03$\pm$0.10 & 19.17$\pm$0.07 & 18.06$\pm$0.05 & 15.45$\pm$0.04 & 14.66$\pm$0.06 & 19.33$\pm$0.02 & 19.14$\pm$0.03 & $-$27.49 & P200/DBSP & 2700 & 2016-09-06\&09\\
SDSS J104836.73+333947.66 & 5.61 & 22.63$\pm$0.17 & 21.30$\pm$0.08 & 19.79$\pm$0.07 & 18.91$\pm$0.10 & 17.03$\pm$0.11 & 16.51$\pm$0.24 & 20.05$\pm$0.04 & 19.86$\pm$0.06 & $-$27.01 & P200/DBSP & 900 & 2017-04-21\\
SDSS J105405.32+463730.25 & 5.47 & 22.68$\pm$0.21 & 20.76$\pm$0.08 & 19.87$\pm$0.12 & 18.53$\pm$0.07 & 16.74$\pm$0.08 & 15.91$\pm$0.13 & 19.80$\pm$0.03 & 19.60$\pm$0.06 & $-$26.97 & P200/DBSP & 900 & 2017-04-16\\
SDSS J121811.19+180750.54 & 5.30 & 22.02$\pm$0.09 & 20.34$\pm$0.04 & 19.83$\pm$0.07 & 19.16$\pm$0.13 & 17.35$\pm$0.15 & 16.71$\pm$0.33 & 20.25$\pm$0.02 & 20.00$\pm$0.04 & $-$26.60 & P200/DBSP & 900 & 2017-04-16\\
SDSS J123606.05+465725.13 & 5.55 & 22.61$\pm$0.16 & 20.48$\pm$0.04 & 19.60$\pm$0.06 & 18.69$\pm$0.13 & 16.45$\pm$0.07 & 15.48$\pm$0.10 & 19.79$\pm$0.02 & 19.75$\pm$0.04 & $-$27.23 & P200/DBSP & 900 & 2017-04-16\\
SDSS J131440.41+543237.21 & 5.52 & 22.45$\pm$0.16 & 20.55$\pm$0.05 & 19.59$\pm$0.07 & 18.80$\pm$0.09 & 16.69$\pm$0.08 & 15.69$\pm$0.10 & 19.85$\pm$0.02 & 19.69$\pm$0.04 & $-$27.21 & P200/DBSP & 900 & 2017-04-16\\
SDSS J150036.84+281603.03 & 5.55 & 22.67$\pm$0.18 & 20.07$\pm$0.03 & 19.06$\pm$0.05 & 18.25$\pm$0.07 & 16.28$\pm$0.05 & 15.50$\pm$0.08 & 19.42$\pm$0.02 & 19.33$\pm$0.03 & $-$27.60 & P200/DBSP & 2360 & 2016-09-05\&11\\
SDSS J161435.36+011444.79 & 5.78 & 23.36$\pm$0.44 & 21.54$\pm$0.16 & 19.68$\pm$0.11 & 19.14$\pm$0.13 & 16.96$\pm$0.11 & 16.42$\pm$0.22 & 19.66$\pm$0.02 & 19.98$\pm$0.06 & $-$26.89 & P200/DBSP & 1200 & 2016-09-06\\
SDSS J165042.26+161721.50 & 5.52 & 22.09$\pm$0.09 & 20.71$\pm$0.04 & 19.28$\pm$0.05 & 18.62$\pm$0.09 & 16.26$\pm$0.06 & 15.40$\pm$0.09 & 19.72$\pm$0.02 & 19.64$\pm$0.04 & $-$27.25 & P200/DBSP & 3300 & 2016-09-06\&07\\
SDSS J231738.25+224409.63 & 5.50 & 21.76$\pm$0.09 & 20.38$\pm$0.04 & 19.16$\pm$0.07 & 18.04$\pm$0.05 & 16.14$\pm$0.06 & 15.32$\pm$0.09 & 19.20$\pm$0.01 & 19.20$\pm$0.03 & $-$27.38 & P200/DBSP & 1800 & 2016-09-05\\
\hline
\end{tabular}
\raggedright {\bf Note.}1. These 15 quasars were all observed using P200/DBSP with grating G316 (R $\sim$ 960 at 7500 \AA\ ) and 1$\farcs$5 slit.\\ 2. These 15 quasars are all from UHS-based selection. So in NIR wavelength, only $J$ band data was used without $H$ and $K$ information.
\end{sidewaystable}

For spectroscopic identifications, we set three priority ranks of our candidate sample. The region that is closer to the color-color cut line has lower priority, considering the facts that there are more candidates in the region closer to star locus and a candidate located closer to star locus has higher probability to be a M dwarf. As shown in Figure 2, there are three subsamples, P1, P2 and P3, with priority from high to low. P1 includes 40 objects, P2 has 47 objects, and P3 has 80 objects. Except for 8 previously known objects, there are 36, 44, and 79 candidates in P1, P2, and P3 respectively. We carried out the optical spectroscopic identifications using the Palomar Hale 5.1m telescope (P200) and the MMT 6.5m telescope in September 2016, and in March, April, October and November 2017. We observed 102 candidates, 80 with P200/DBSP and 22 with MMT/Red Channel, and discovered 15 new quasars including 13 $z \sim 5.5$ ($5.3 \le z \le 5.7$) quasars. All information regarding photometry and spectroscopic identification for these 15 new quasars is listed in Table 2.

With P200/DBSP, we used the grating G316 (R $\sim$ 960 at 7500\AA) and 1$\farcs$5 slit. The DBSP RED side covers wavelength range from $6050 \AA\ $ to $10000 \AA\ $. When we were using MMT/Red Channel spectrograph \citep{schmidt89}, we chose the 270 $\rm l  mm^{-1}$ grating with a central wavelength at 7900 $\rm \AA$ (8500 $\rm \AA$), which provided spectral coverage from $\sim$ 6100 to 9700 $\rm \AA$ ($\sim$ 6700 to 10300 $\rm \AA$). Based on the seeing conditions, we used the $1\farcs0$ or $1\farcs5$ slit, providing resolutions of $R\sim 640$ and $R\sim 430$, respectively. All P200 and MMT spectra were reduced using standard IRAF routines. Spectra of 15 new quasars are presented in Figure 3.

\begin{figure*}%f3
\centering
\epsscale{1.3}
\plotone{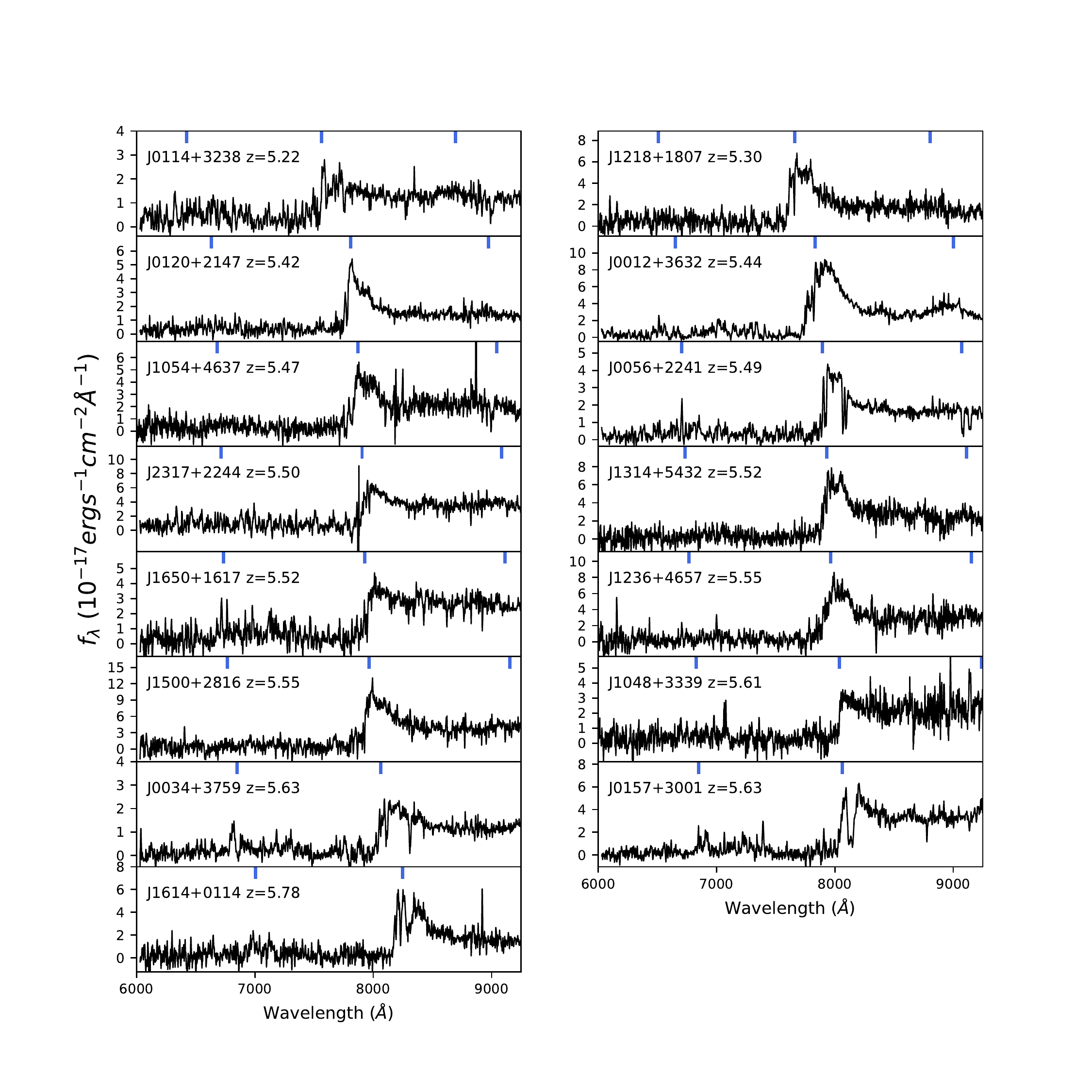}
\caption{The spectra of 15 newly discovered quasars. The blue vertical lines denote the $\rm Ly \beta$, $\rm Ly \alpha$ and Si\,{\sc iv} emission lines. All spectra are smoothed with a 3 pixel boxcar. All spectra are corrected for Galactic extinction using the \cite{cardelli89} Milky Way reddening law and E(B $-$ V) derived from the \cite{schlegel98} dust map.}
\end{figure*}

We measure the quasar redshifts by visually matching the observed spectrum to a quasar template using an eye-recognition assistant for quasar spectra software \cite[ASERA;][]{yuan13}. The matching is based on a series of emission line, $\rm Ly \beta$, $\rm Ly \alpha$, N\,{\sc v}, O\,{\sc i}/Si\,{\sc ii}, and Si\,{\sc iv}. The typical uncertainty of the redshift measurement is around 0.03. We do not include the systematic offset of Ly$\alpha$ emission line \citep[e.g.,][]{shen07}, which is typical $\sim$ 500 km/s and much smaller than the uncertainty of matching. We fit a power-law continuum for each observed spectrum to calculate $M_{1450}$, the absolute magnitude at rest-frame 1450\AA. 
We normalize the power-law continuum to match the visually identified continuum windows which contain minimal contribution from quasar emission lines and sky OH lines. All quasar spectra used are scaled by their SDSS $i$-band magnitude as our spectra do not fully cover SDSS $z$-band. The uncertainties of power-law continuum fitting are much smaller than the photometric errors, therefore the uncertainties of $M_{1450}$ are comparable to SDSS $i$-band photometric errors. The redshifts and $M_{1450}$ of these 15 new quasars are also listed in Table 2.

\subsection{Combined Result of $z \sim 5.5$ Quasar Survey}

The observations presented here in combination with the results in Paper I complete our SDSS--ULAS/VHS/UHS--WISE based $z \sim 5.5$ quasar survey. In total we observed 130 candidates out of the 167 candidates in the final candidate sample. There are 7 known quasars and one galaxy from other works, and 29 objects without spectroscopic observation.
Thirty two P1, forty two P2 and fifty six P3 candidates were observed, among which 37 new quasars at $4.5 \le z \le 5.78$ have been discovered. 
The quasar fraction and spectroscopic completeness in P1, P2 and P3 ranks will be discussed in Section 4.2.
In our survey, we identified 31 new $z \sim 5.5$ quasars ($5.3 \le z \le 5.7$) in total, and 30 quasars are in the final sample\footnote{Quasars, J0957+1016 ($z = 5.14$), J2225+0330 ($z = 5.24$) and J1006--0310 ($z = 5.5$) were discovered in Paper I but do not meet the $zyJ$ cut. Therefore, they are not included in the final sample.}. 
Our $z \sim 5.5$ quasar selection method yields a total success rate of 23.1\% (30 of 130): it is 28.6\% in ULAS/VHS based sample (16 of 56) and 20.0\% in UHS based sample (15 of 75, one overlap between two sub-samples). The ULAS/VHS selection has a higher success rate since $H$ and $K$ bands are included in selection. 

At $5.3 \le z \le 5.7$, there are 32 previously known quasars. Among them, 6 quasars are in our survey magnitude limit and have required photometric data. Among these 6 quasars, four were rejected by $i-z$ cut and one was rejected by $K-$W2 cut. Only one is included in the final sample, which is a SDSS quasar at $z = 5.31$.
Together with this one known quasar, there are 31 $z \sim 5.5$ quasars ($5.3 \le z \le 5.7$) in the final sample of 167 candidates, except for 29 unobserved candidates. 
The four quasars rejected by $i-z > 0.5$ cut are from previous $z \sim 5$ quasar surveys which focused on the bottom-right region in the $riz$ color space (see Figure 1), and thus are biased when used to qualify the selection completeness. 
We will estimate our selection function with a sample of simulated quasars in Section 4.2. In Figure 4, we plot the redshift distribution of our new quasars, compared with all previously known quasars with $z <$ 20.5 mag and all SDSS--ULAS/VHS/UHS--ALLWISE detected known quasars within the magnitude limit of our survey, which is $z <$ 20.5 mag for ULAS/VHS area, and $z <$ 20.0 mag for UHS area. Our survey has doubled the number of known quasars at $z \sim$ 5.5 within our magnitude limit.

\begin{figure}%f4
\centering
\epsscale{1.35}
\plotone{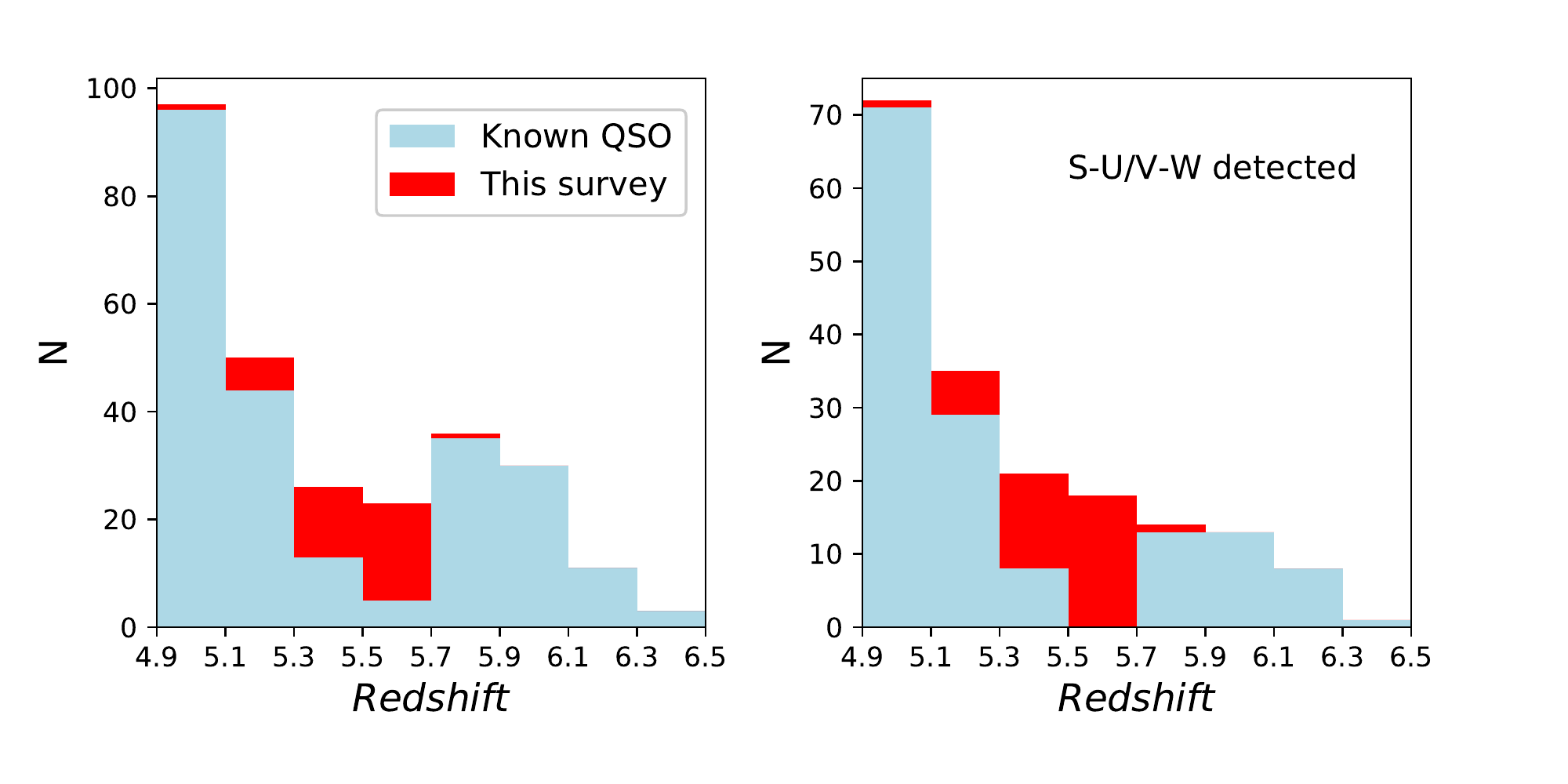}
\caption{ $Left:$ The redshift distribution of newly discovered quasars from our survey (red), compared with that of all previously known quasars at $z >$ 4.9 with $z$ band magnitude brighter than 20.5 mag. Here we only include quasars within the flux limit of our survey; there are also 11 previously known $z \sim 5.5$ quasars with z band magnitude fainter than 20.5. 
As shown, there is an obvious gap at 5.1 $< z <$ 5.7, especially at $z \sim$ 5.5.
We use the known quasar sample from the combination of the known quasar catalogs in \cite{wang16}, \cite{banados16} and \cite{jiang16}. 
Using the SDSS--PS1--ULAS/VHS/UHS--WISE color-color selection, we have discovered 31 new quasars at 5.3 $< z <$ 5.7 and 9 lower/higher redshift quasars. Our optical NIR color selection is effective for searching quasars located in this redshift gap. 
$Right:$ The comparison between new quasars and all SDSS--ULAS/VHS/UHS--ALLWISE (S-U/V-W) detected known quasars in our survey magnitude limit, $z <$ 20.5 for ULAS/VHS, and $z <$ 20.0 for UHS area. Our survey has filled the redshift gap within our survey area and magnitude limit.
}
\end{figure}

\section{Quasar Spatial Density at $z \sim 5.5$}
\subsection{Sample Construction}

From the $z\sim5.5$ quasars presented in \S2.4, we define a uniformly selected, flux-limited subsample, which enables us to calculate the quasar number density at $z \sim 5.5$ for the first time. The complete quasar sample is defined as follows:
\begin{itemize}
\item Quasars in the redshift range of $5.2 \le z \le 5.7$. As shown in the Figure 5, our selection function shows a relative high completeness at $z \sim 5.2-5.3$ due to the slow evolution of quasar's color at this redshift range. So we include this redshift range (6 quasars) for number density measurement. 
\item Quasars in the SDSS--ULAS or SDSS--UHS survey footprint. We exclude quasars from the SDSS--VHS footprint, considering the fact that it is difficult to count the overlapped sky coverage between SDSS and released VHS and there are only few quasars in this area. Therefore, we decide to remove five VHS quasars for the number density calculation.
\item Quasars in the magnitude range of $M_{1450} < -26.4$ in the UHS area and $M_{1450} < -26.2$ in the ULAS area. In the UHS selection (see Figure 5), simulated quasars at $z = 5.7$ with $M_{1450} = -26.4$ hit the survey limit, SDSS $z$ band apparent magnitude $= 20$ mag. One UHS quasar is rejected in this step. In the ULAS sample, the lowest luminous quasar has $M_{1450} = -26.20$. So we choose a limit of $M_{1450} = -26.20$ for the ULAS sample.
\end{itemize}
Based on these three criteria, we obtain a sample including 31 quasars, as listed in Table 3. There are 14 ULAS quasars and 18 UHS quasars (one overlap). We also list which subsample (P1, P2 or P3) each quasar belongs to, which will be used for the spectroscopic incompleteness correction. All $M_{1450}$ are measured using the method described above, including the SDSS quasar.

\begin{deluxetable*}{l c c c c c c c }%T3
\tablecaption{Quasar sample for number density calculation. \label{tbl-3}}
\tablewidth{0pt}
\tablehead{
\colhead{Name} & \colhead{Redshift} &\colhead{SDSS$\_z$}  &\colhead{SDSS$\_z$err} &
\colhead{$M_{1450}$}&
\colhead{NIRdata}  & 
\colhead{Reference}&  \colhead{Priority} 
}
\startdata
  SDSS J001232.88+363216.1 & 5.44 & 19.30 & 0.05 & -27.25& UHS & Thiswork & P1\\
  SDSS J003414.35+375954.0 & 5.63 & 19.99 & 0.09 & -26.44 & UHS & Thiswork & P1\\
  SDSS J005656.04+224112.2 & 5.49 & 19.89 & 0.10 & -26.77 & UHS & Thiswork & P3\\
  SDSS J010806.60+071120.6 & 5.53 & 19.57 & 0.09 & -27.12 & ULAS & Paper I & P2\\
  SDSS J012053.93+214706.2 & 5.42 & 19.95 & 0.12 & -26.49 & UHS & Thiswork & P1\\
  SDSS J015533.28+041506.7 & 5.37 & 19.26 & 0.06 & -27.03 & ULAS & Paper I & P2\\
  SDSS J015745.45+300110.7 & 5.63 & 19.17 & 0.07 & -27.49 & UHS & Thiswork & P3\\
  SDSS J082933.10+250645.6 & 5.35 & 19.67 & 0.08 & -26.91 & ULAS & Paper I & P2\\
  SDSS J101637.71+254131.9 & 5.64 & 18.96 & 0.05 & -27.74 & UHS & Paper I & P1\\
  SDSS J102201.91+080122.2 & 5.30 & 19.12 & 0.05 & -27.56 & ULAS & Paper I & P2\\
  SDSS J104836.73+333947.7 & 5.61 & 19.79 & 0.07 & -27.01 & UHS & Thiswork & P2\\
  SDSS J105405.32+463730.3 & 5.47 & 19.87 & 0.12 & -26.97 & UHS & Thiswork & P3\\
  SDSS J113308.78+160355.7 & 5.61 & 19.71 & 0.12 & -27.42 & ULAS\&UHS & Paper I & P1\\
  SDSS J113414.23+082853.3 & 5.69 & 20.31 & 0.13 & -26.34 & ULAS & Paper I & P1\\
  SDSS J114706.41$-$010958.2 & 5.31 & 19.23 & 0.04 & -27.37 & ULAS & Paper I & P3\\
  SDSS J114946.45+074850.6 & 5.66 & 20.30 & 0.14 & -26.33 & ULAS & Paper I & P1\\
  SDSS J120207.78+323538.8 & 5.26 & 18.42 & 0.05 & -28.04 & UHS & SDSSDR12 & P2\\
  SDSS J121811.19+180750.5 & 5.30 & 19.83 & 0.07 & -26.60 & UHS & Thiswork & P2\\
  SDSS J123606.05+465725.1 & 5.55 & 19.60 & 0.06 & -27.23 & UHS & Thiswork & P1\\
  SDSS J131440.41+543237.2 & 5.52 & 19.59 & 0.07 & -27.21 & UHS & Thiswork & P1\\
  SDSS J131720.78$-$023913.0 & 5.25 & 20.08 & 0.14 & -26.20 & ULAS & Paper I & P2\\
  SDSS J133556.24$-$032838.2 & 5.67 & 18.89 & 0.04 & -27.69 & ULAS & Paper I & P2\\
  SDSS J143605.00+213239.2 & 5.22 & 19.28 & 0.06 & -27.04 & UHS & SDSSDR12 & P1\\
  SDSS J150036.84+281603.0 & 5.55 & 19.06 & 0.05 & -27.60 & UHS & Thiswork & P2\\
  SDSS J151339.64+085406.5 & 5.47 & 19.89 & 0.09& -26.74 & ULAS & Paper I & P1\\
  SDSS J152712.86+064121.9 & 5.57 & 19.95 & 0.10 & -26.85 & ULAS & Paper I & P2\\
  SDSS J165042.26+161721.5 & 5.52 & 19.28 & 0.05 & -27.25 & UHS & Thiswork & P1\\
  SDSS J165902.12+270935.2 & 5.31 & 18.70 & 0.04 & -27.85 & UHS & SDSSDR7 & P1\\
  SDSS J231738.25+224409.6 & 5.50 & 19.16 & 0.07 & -27.38 & UHS & Thiswork & P2\\
  SDSS J233008.71+095743.7 & 5.30 & 19.78 & 0.10 & -26.68 & ULAS & Paper I & P3\\
  SDSS J235824.04+063437.4 & 5.32 & 19.54 & 0.08 & -27.19 & ULAS & Paper I & P1\\
 \enddata
\tablecomments{1. Known SDSS quasars are from SDSS DR7 \& DR12 quasar catalog \citep{schneider10, paris16}. 2. The quasar selected by both ULAS and UHS is treated as a ULAS quasar for the number density calculation. The overlapped region is also be subtracted from UHS area. }
\end{deluxetable*}

\subsection{Survey Completeness}

\begin{figure}%f5
\centering
\epsscale{1.2}
\plotone{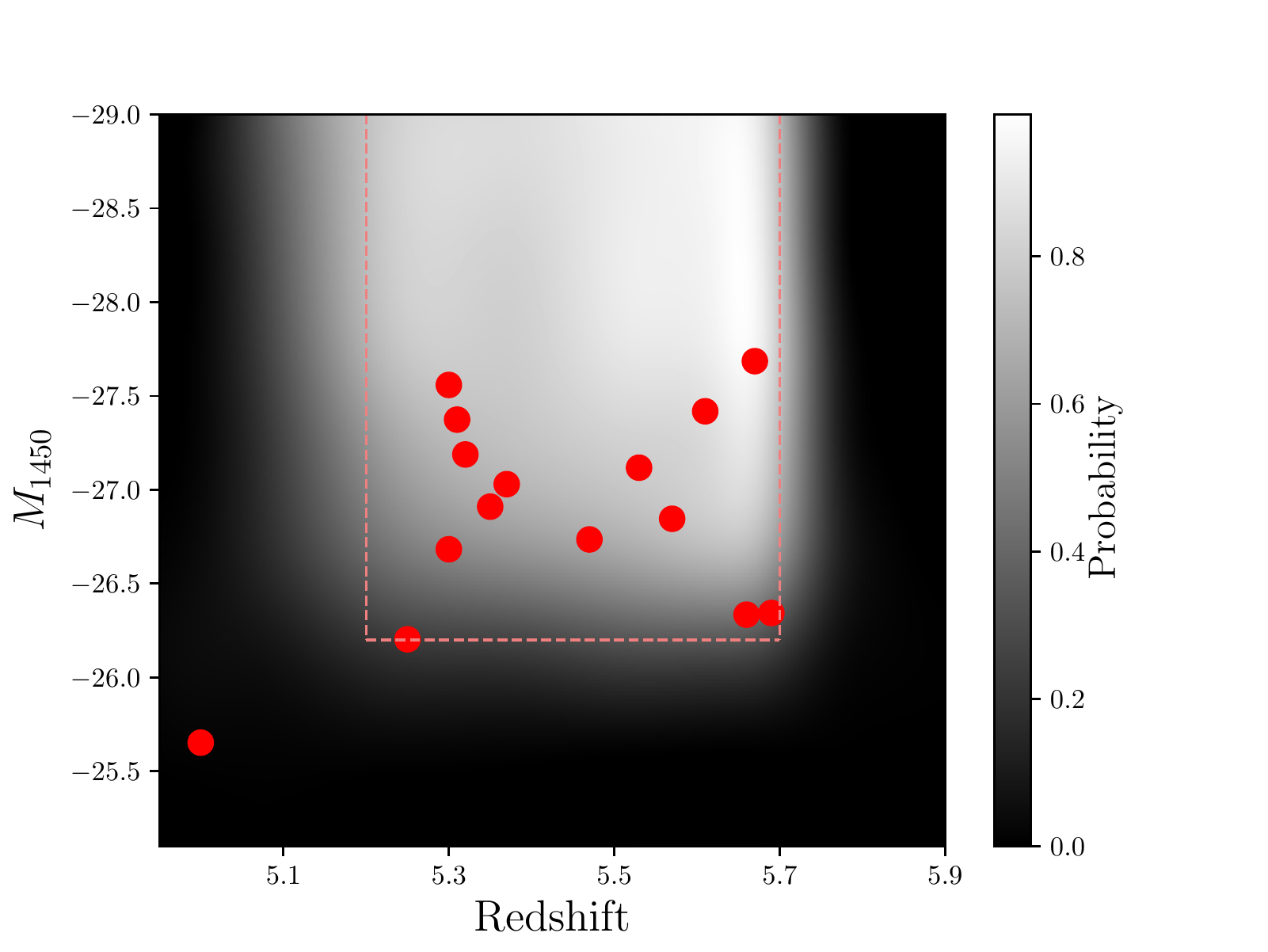}
\plotone{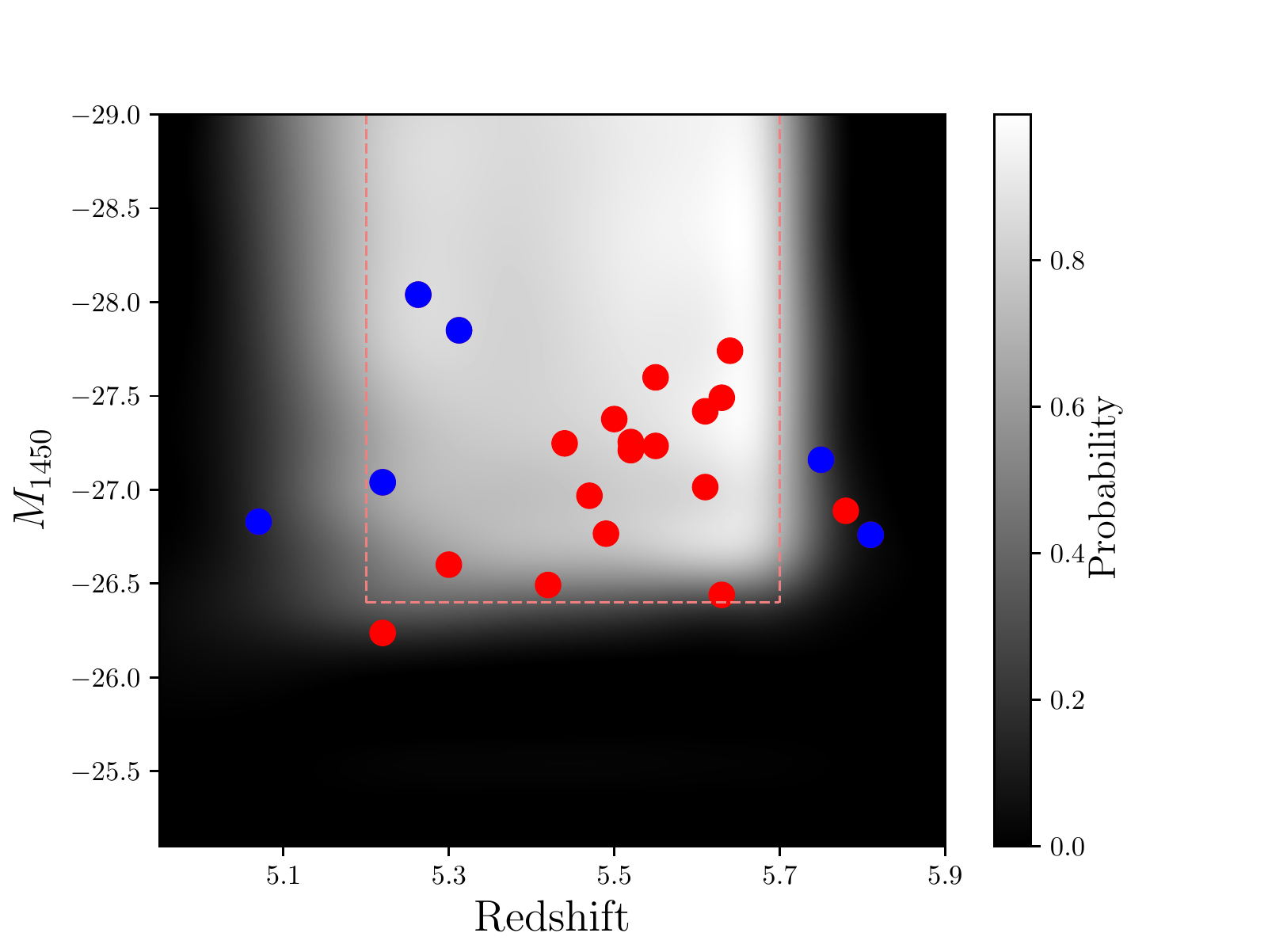}
\caption{The selection function of SDSS--PS1--ULAS--WISE ({\it Top}) and SDSS--PS1--UHS--WISE ({\it Bottom}) color selections. Red points represent quasars from our survey, and blue points show previously known quasars selected by our method. They are from SDSS DR7 \& DR12 \citep{schneider10, paris16}, \cite{banados16} and \cite{jiang16}. 
The mean selection probability at $5.2 < z < 5.7$ and $-28 <M_{1450} < -26.4$ is $\sim$ 73.4\% in the ULAS area, and $\sim$ 77.9\% in the UHS area. 
%Because in UHS area, there is no $H$ and $K$ colors included in the selection criteria.
Towards the fainter end, the decreasing of selection probability is caused by the increasing photometric uncertainties( especially in W2 band) and the survey magnitude limit. The red dashed lines bound the quasar sample used for number density calculation.}
\end{figure}

In this subsection, we discuss the completeness of our selection method and spectroscopy, which will be used for the incompleteness correction for the number density calculation. For our visually image checking procedure, the incompleteness is $\sim$ 2\% -- 4\%, which was estimated by \cite{yang16} using a sample of randomly selected SDSS images of point sources. It is difficult to obtain a more accurate value of this incompleteness and this effect is much smaller than the error of number density calculation, thus image selection is not included in incompleteness correction.

% WISE incompleteness, Pan-STARR incompleteness
To calculate the completeness of our color selection criteria, we generate a sample of simulated quasars using the quasar model from \cite{mcgreer13}. We extend this model toward redder wavelengths to cover the ALLWISE W1, W2 bands for quasars at $z$ = 5 to 6 \citep{yang16}. Based on this model, a total of $\sim$ 200,000 simulated quasars have been generated and evenly distributed in the ($M_{1450}$, $z$) space of 5$\le z \le$6 and $-30 < M_{1450} < -25$. We assign optical photometric errors using the magnitude-error relations from the SDSS main survey. For $J,H,K$ bands, we use the ULAS photometric errors. Photometric data of the UHS survey is from the same instrument and filter, and has similar depth to ULAS. Therefore, we use the same magnitude-error relation for UHS $J$.

%selection function
The ALLWISE detection depth is highly dependent on the sky position, which will affect the detection incompleteness and photometric uncertainties. We model the coverage-dependent detection incompleteness and photometric uncertainties of ALLWISE using the ALLWISE coverage map within the SDSS--ULAS/VHS area. We follow the procedure outlined in \cite{yang16}.
The PS1 depth is also not homogeneous. As we are using PS1 for bright objects (SDSS $z$ band magnitude brighter than 20.0 mag), the detection incompleteness will only slightly affect our selection function. When we matched our candidate sample with PS1 DR1 catalog, only $< 2$\% SDSS sources do not have PS1 detections. Therefore, we only model the coverage-dependent photometric uncertainties of PS1 data. We build the PS1 depth map in the SDSS-ULAS and SDSS-UHS area respectively, and apply the same method that is used for ALLWISE to model the PS1 photometric uncertainties. 

We calculate the fraction of simulated quasars selected by our selection pipeline in each magnitude-redshift bin ($\Delta$M = 0.1 and $\Delta z$ = 0.05).
The selection function is a function of redshift and absolute magnitude $M_{1450}$, as shown in Figure 5. We find a 73\% mean selection completeness of SDSS--PS1--ULAS--{\it WISE} selection, and 78\% in UHS based selection. The difference is caused by the $H \& K$ colors used in ULAS area.
The highly complete region extends to $z \sim 5.2$, which is the result of the slow evolution of quasars in $r-i$ and $i-z$ colors from $z=$ 5.1 to 5.3 (see Figure 1). In order to improve the completeness of $z \sim 5.5$ quasar selection, we use a relatively relaxed $riz$ cut, and thus can include some lower redshift quasars, which is also shown in the redshift distribution in Figure 3. At the high redshift end, the $i-z < 2.2$ cut will restrict the selected sample to $z<5.7$, due to the rapid increasing of quasar $i-z$ color from $z =$ 5.7 to 5.8, which is represented by a sharp edge at $z \sim 5.7$.

%spectroscopic incompleteness
We estimate the spectroscopic incompleteness by assuming that the fraction of $z \sim 5.5$ quasars in the observed candidates and unobserved candidates are the same in each priority sub-sample. In P1, there are 40 candidates (including 4 previously known objects). We observed 32 candidates, and obtained 14 $\sim 5.5$ new quasars. In P2, there are 47 candidates (including 3 known objects); among them 42 candidates are observed by our survey, from which we got 11 new $z \sim 5.5$ quasars. In P3, there are 80 candidates (including 1 known object); among them 56 candidates are observed, and five are new $z \sim 5.5$ quasars. Therefore, the success rate at P1, P2 and P3 is 43.8\% (14/32), 26.2\% (11/42) and 8.9\% (5/56), respectively. The spectroscopy completeness is 88.9\% (32/36), 95.5\% (42/44) and 70.9\% (56/79) in P1, P2 and P3 respectively.

\subsection{Spatial Density and Its Evolution}

\begin{figure}%f6
\centering
\epsscale{1.3}
\plotone{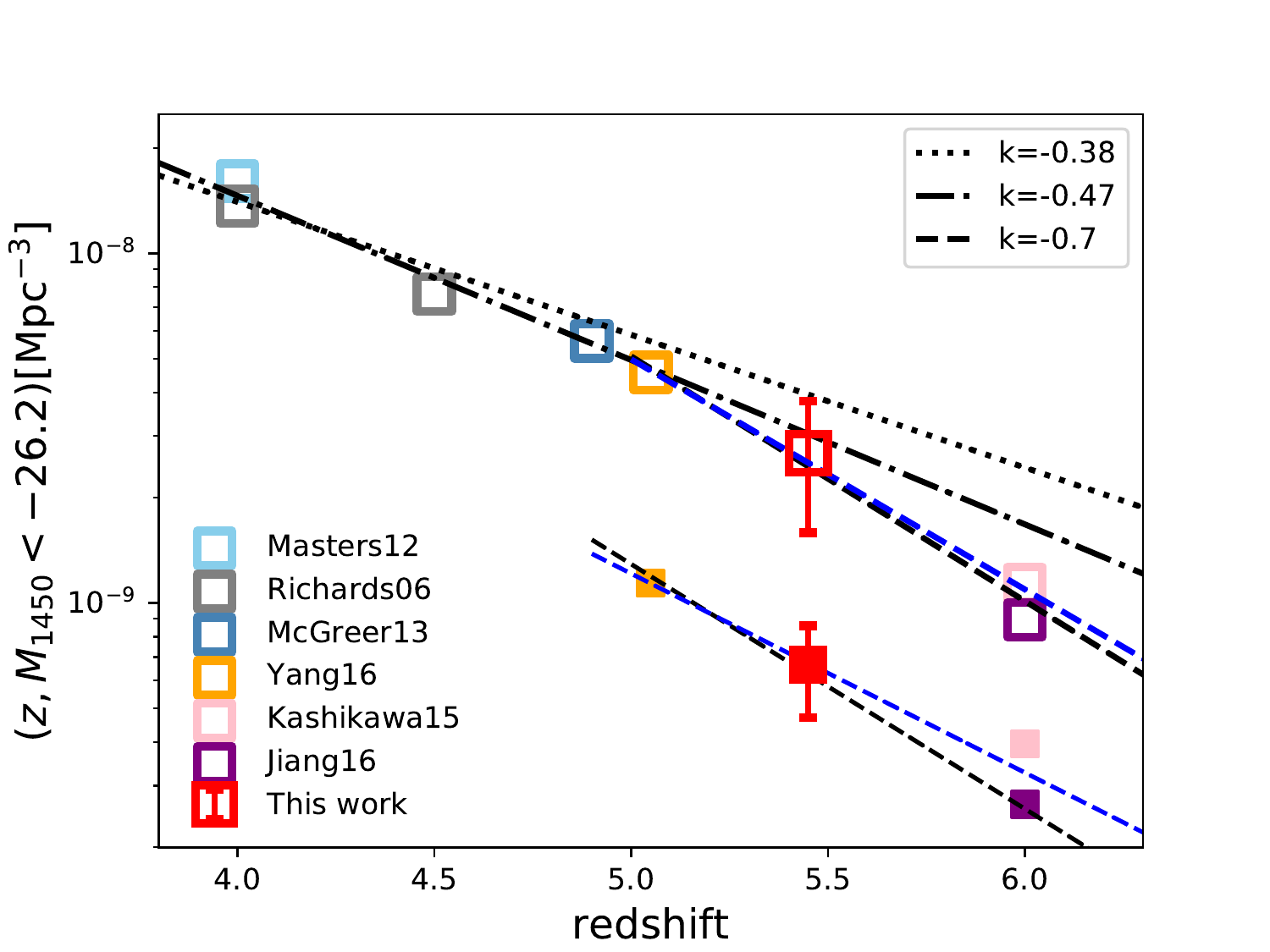}
\caption{Evolution of quasar spatial density at $M_{1450} < -26.2 (-27)$ at high redshift. All squares without error bars are spatial density based on quasar luminosity functions in previous works. The spatial densities at $z = 4$ and $z =4.5$ are from \cite{masters12} (light blue open square) and \cite{richards06} (grey open squares). The points at $z = 4.9$ and 5.05 are u from \cite{mcgreer13} (steel blue) and \cite{yang16} (orange). At $z=6$, we are using the results from \cite{kashikawa15} (pink) and \cite{jiang16} (purple). The red squares are results from our quasar sample. All open squares represent the density at $M_{1450} < -26.2$, and small solid squares denote the density at $M_{1450} < -27$. The black dotted, dash-dotted and dashed lines represent the evolution with different slopes. The two blue dashed lines are our fits of $z > 5$ points within $M_{1450} < -26.2$ and $-27$, with $k= -0.66 \pm 0.05$ and $k=-0.57 \pm 0.06$, respectively. Our result shows good agreement with the rapid decline of quasar spatial density at $z > 5$.}
\end{figure}

Our sample is located at the bright end of the quasar population, with comparable or brighter luminosity than the break magnitude which is expected to be in the range of $M_{1450}^{*} \sim -25 - -27$ based on the recent QLFs at $z \sim 5$ and 6 \citep{mcgreer13,yang16,jiang16,kashikawa15}. In addition, at $z \sim 5.5$, there is no previous constraint on the the faint end. So using only our new sample can not provide reasonable measurement of the double power law QLF at $z \sim 5.5$. Our survey provides a uniform quasar sample for the measurement of quasar spatial density at $z \sim 5.5$ for the first time, which will place a new constraint on the density evolution from redshift 6 to 5.
We calculate the quasar number density at $5.2 \le z \le 5.7$ in the magnitude range of $M_{1450} < -26.2$ using the quasar sample discussed above and applying all the incompleteness estimates. The spatial density is derived from the $1/V_{a}$ method, as the following equation.
\begin{equation}
	\rho = \sum_i\frac{1}{V_a^i} ~,~~ 
	\sigma(\rho)=
	  \left[\sum_i\left(\frac{1}{V_a^i}\right)^2\right]^{1/2} ~,
\end{equation}
where the sum is over all quasars more luminous than the minimum luminosity $M$.
We estimate the effective survey area of SDSS-ULAS/UHS using the Hierarchical Equal Area isoLatitude Pixelization \cite[HEALPix][]{gorski05}, and obtain $\sim$ 3520 $\rm deg^{2}$ in the ULAS area and $\sim$ 6780 $\rm deg^{2}$ in the UHS area.

The spatial density is shown in Figure 6, compared with densities at $M_{1450} < -26.2$ at different redshifts derived from integrating quasar luminosity functions in \cite{richards06}, \cite{masters12}, \cite{mcgreer13}, \cite{yang16}, \cite{kashikawa15} and \cite{jiang16}. Our spatial density shows good agreement with the predicted number from recent QLFs at redshift $z \sim 5$ and 6 \citep{mcgreer13, yang16, jiang16, kashikawa15, kulkarni18}. 
\cite{fan01b} fit an exponential decline to the space density at high redshifts and find that $\rho(M_{1450} < -25.5, z) \sim 10^{kz}$, where $k = -0.47$ at $z>3.6$, corresponding to a factor of three per unit redshift. \cite{mcgreer13} fit the density at $z =4.25, 4.9$ and 6 and suggest a $k= -0.38$ at redshift from 4 to 5 but a steeper slope between $z = 5$ and 6, with a $k = -0.7$. The rapid decline has also been suggested by \cite{jiang16}. We fit the points at $z = 5.05$, 5.45 and 6, and find the $k= -0.66 \pm 0.05$. The result is highly consistent with the rapid decline from $z =5$ to $z =6$ suggested by \cite{mcgreer13, jiang16}. We also calculate the spatial density at more luminous end, $M_{1450} < -27$: the decline is slower with $k=-0.57 \pm 0.06$, which is also consistent with the result from \cite{mcgreer13}. Although the uncertainties of $k$ prevent detailed study of the evolution model, our result suggests that quasar spatial density decline more rapidly at $z =5$ to 6 than at lower redshifts and the decline is stronger for the fainter quasars.

\section{PS1 Selection}

Our SDSS--ULAS/UHS/VHS--{\it WISE} survey shows that the optical-infrared color-color method is effective for $z \sim 5.5$ quasar selection. The SDSS-based optical colors limit the survey area and depth. After the SDSS-based selection is completed, we further extend the selection to fainter flux and wider area, beyond the limit of SDSS photometry. 
As discussed above, the PS1 survey covers 3$\pi$ sky with $g,r,i,z,y$ bands, providing a larger survey area. PS1 is also deeper than SDSS, offering excellent dataset for searching fainter quasars. 
In this Section, we apply our $z \sim 5.5$ quasar selection method and the $z \sim 5$ quasar selection procedure from \cite{wang16} in the new survey in the whole PS1 area for searching $z \sim 5 $ and 5.5 quasars. 
In this survey, we only use optical data from PS1.
The PS1 $g,r$ and $i$ bands are similar to the SDSS bands, and the PS1 $z$ band is bluer and narrower. 
Thus we can also use PS1 data for the $riz$ color-color cut but do need to modify the criteria according to the PS1 colors. The PS1 $y$ band is helpful to improve the efficiency of quasars selection, as discussed in Section 2. We use the PSFMag - KronMag $<$ 0.3 to separate quasars from galaxies. 
In the infrared, we use WISE W1 and W2 and an additional $J$ band is used for $z \sim 5.5$ quasar selection only.
In this Section, we use the $g, r, i,z$ and $y$ photometry from  PS1 DR1, after Galactic extinction correction following \cite{schlafly11}.

\subsection{Color Selections}
\subsubsection{$z \sim 5$ Quasar Selection}

\begin{figure}%f7
\centering
\epsscale{1.0}
\plotone{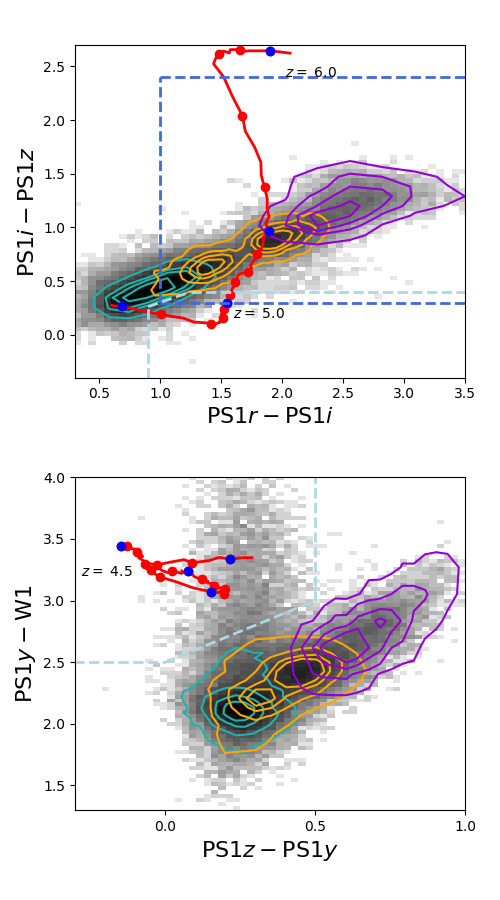} 
\caption{The PS1 $r-i$/$i-z$ color-color diagram ({\it Top}) and PS1$z-$ PS1$y$/PS1$y-$W1 color-color diagram ({\it Bottom}). The red solid lines with red dots are color tracks generated by calculating mean colors of simulated quasars from $z = 4.5 $ to $z = 6$. The color track has a step of $\Delta z$ = 0.1. Blue dots denote colors at $z = 4.5, 5.0, 5.5$ and 6. Green, orange and purple contours show the locus of M1--M3, M4--M6 and M7--M9 type M dwarfs. Contours in the $zy$W1 plot only show $riz$ selected M dwarfs.
The light blue dashed lines are selection criteria of $z \sim 5$ quasar candidate sample. 
The blue dashed lines in $riz$ plot represent $riz$ cut used for $z \sim 5.5$ sample.
Again, for $z \sim 5.5$ quasar selection, the PS1 $riz$ selection is used to exclude early type dwarfs and to restrict quasars to be at $z \sim 5.5$, while it also includes most of M4--M9 dwarfs.
} 
\end{figure}

For the $z \sim 5$ quasar selection, we follow our previous SDSS-WISE selection method in \cite{wang16}, starting with the PS1 $riz$ color-color diagram. The criteria are modified according to PS1 filters, as shown in Figure 7. 
In \cite{wang16}, we used $W1-W2>0.5$ in order to reject as many dwarfs as possible. However, this color cut will reject more than half of quasars between $z\sim4.5$ and $z\sim4.8$. In order to maintain higher completeness at this redshift range, we relax our {\it WISE} color cuts but add additional PS1$z-$ PS1$y/$PS1$y-$w1 color cuts at the same time to reduce contaminations. We limit the galactic latitude to be $|$b$| > 15 ^{\circ}$.
The detailed selection criteria are described as following, and also in Figure 7.

\begin{equation}
S/N(g)<5.0 ~or~ g < 0 ~or~ g-r>1.5
\end{equation}
\begin{equation}
r-i>0.9 ~and~  i-z<0.4
\end{equation}
\begin{equation}
0.4\times(r-i) - 0.1 > (i-z)
\end{equation}
\begin{equation}
y-(W1+2.699)>-0.2 ~and~ z-y<0.5
\end{equation}
\begin{equation}
y-(W1+2.699)>(z-y)-0.2
\end{equation}
\begin{equation}
(W1+2.699)-(W2+3.339)>-0.3
\end{equation}

\subsubsection{$z \sim 5.5$ Quasar Selection}
For $z \sim 5.5 $ quasars, we follow the method used in the UHS area, using $riz$, $zyJ$ and $J$W1W2 colors. 
We started with PS1 $riz$ cuts listed as following (also see Figure 7),

\begin{equation}
S/N(g)<3 ~or~ g < 0 ~or~ g-r>1.8
\end{equation}
\begin{equation}
S/N(i) > 3 ~and~ S/N(z) > 3
\end{equation}
\begin{equation}
r-i>1.0 ~or~ S/N(r)<3
\end{equation}
\begin{equation}
 0.3 < i-z < 2.4 
\end{equation}

and then cross match with ULAS, UHS, UKIDSS - Galactic Clusters Survey (GCS), VHS and VISTA Kilo-Degree Infrared Galaxy (VIKING) surveys \cite[VIKING][]{edge13} for the $J$ band photometric data. The overlap areas between PS1 and VHS/VIKING provide new optical/NIR covered area for quasar selection in the Southern sky. We still use the ALLWISE photometric data. The $zyJ$ and $J$W1W2 selection are the same to equations (6) - (9) (also Figure 1), and (12) \& (13) (also Figure 2), but we relax equation (13) to
\begin{equation}
 \begin{array}{l}
y-J < 0.8 \\
~or~\\
y-J < -1.5\times(z - y) + 1.5
 \end{array}
\end{equation}
To explore quasar selection at low galactic latitude, in this sample, we limit the galactic latitude to be $|$b$| > 10 ^{\circ}$. Our survey does discover six quasars at low galactic latitude, $10 ^{\circ} < |$b$| < 17 ^{\circ}$, which suggests the feasibility of our method to search for low galactic latitude quasars. 

\subsection{Quasar Identification}
We observed $\sim$ 70 bright candidates, $\sim$ 40 from $z \sim 5$ sample and $\sim$ 30 from $z \sim 5.5$ sample, as backup targets of spectroscopy observing runs with ANU 2.3 m telescope, Lijiang 2.4m telescope, MMT 6.5m telescope and Magellan/Clay 6.5m telescope from 2015 to 2018. We chose candidates based on their brightness, colors, positions and the weather condition during observations. So at this point,  the quasars discovered from this survey do not yet form a complete sample.
For the $z \sim 5$ candidates, objects with $z$ band magnitude brighter than 19.5 usually have high priority. For the $z \sim 5.5$ candidates, we observed candidates with $i-z > 0.4$ and $z < 20$ at high priority. 
We discovered 38 quasars at $4.61 \le z \le 5.07$ from the $z \sim $5 quasar candidate sample and 13 quasars at $4.96 \le z \le 5.71 $ from the $z \sim 5.5$ quasar candidate sample. 
Photometric information of the 51 new quasars are provided in Table 4. 
The observation information and spectra of these new quasars are presented in Table 5 and Figure 8.
The redshifts are measured with the same method in Section 3.1. Due to low signal-to-noise ratios, the redshift uncertainties of broad absorption line quasars and weak line quasars will be larger than 0.03 used in Section 3.1.
Since half of them only have low signal-to-noise ratio spectra, we estimate their $M_{1450}$ using quasar template \citep{selsing16} scaled by PS1 $y$ photometric data of each quasar. 

For spectroscopy on the ANU 2.3m Telescope at Siding Spring Observatory, we used the Wide Field Spectrograph \citep[WiFeS;][]{dopita07,dopita10}, an integral-field double-beam image-slicing spectrograph. We chose R3000 grating on WiFeS which gives a resolution of $R=3000$ at wavelengths between 5300$\rm \AA$ and 9800 $\rm \AA$. 
With the Lijiang 2.4m telescope, we used Grism 5 (G5) of the Yunnan Faint Object Spectrograph and Camera (YFOSC), with dispersion of 185 $\rm \AA$/mm and wavelength coverage from 5000 to 9800 $\rm \AA$. We used a $1\farcs8$ slit which yields a resolution of $R\sim550$. 
With the MMT, we used Red Channel spectrograph \citep{schmidt89} with the 270 $\rm l  mm^{-1}$ grating. 
We also used Magellan Clay/LDSS3 with VPH-Red grism and $1\farcs0$/$1\farcs25$\_center slits, which reach R $\sim$1357 and R $\sim$1086, respectively. All spectra were reduced by IRAF.

Among these 51 new quasars, there are three quasars have detections from the Two Micron All Sky Survey \cite[2MASS;][]{skrutskie06}, J091655--251145, J111054--301129 and J130031--282931. Quasar J091655--251145 and J111054--301129 are the two most luminous quasars among our new discoveries, with $M_{1450} < -29$.

\begin{deluxetable*}{l c c c c c c c c c c}%T4
\tablecaption{Photometric Information of 51 Newly Identified Quasars. \label{tbl-4}}
\tabletypesize{\scriptsize}
\tablewidth{0pt}
\tablehead{
\colhead{Name} & 
\colhead{Redshift} &
\colhead{$M_{1450}$}&
\colhead{PS1\_$r$}  &\colhead{PS1\_$i$} &
\colhead{PS1\_$z$}  &\colhead{PS1\_$y$} &
\colhead{$J$\tablenotemark{a}}  & \colhead{NIR} &
\colhead{W1}  & \colhead{W2} 
}
\startdata
PS1 J001749.00+271052.10 & 4.76 & -27.18 & 21.29$\pm$0.03 & 19.22$\pm$0.01 & 19.27$\pm$0.01 & 18.93$\pm$0.02 & -- & -- & 15.87$\pm$0.05 & 15.43$\pm$0.10\\
PS1 J005243.45+455632.03 & 5.31 & -27.05 & 21.46$\pm$0.05 & 19.92$\pm$0.01 & 19.34$\pm$0.02 & 19.20$\pm$0.05 & 18.37$\pm$0.08 & UHS &16.43$\pm$0.06 & 15.67$\pm$0.11\\
PS1 J013901.97+412358.00 & 4.64 & -27.19 & 20.19$\pm$0.02 & 18.82$\pm$0.01 & 19.01$\pm$0.02 & 18.87$\pm$0.03 & -- & -- & 15.77$\pm$0.05 & 15.13$\pm$0.08\\
PS1 J020523.76+202347.55\tablenotemark{b}  & 4.69 & -27.23 & 20.09$\pm$0.03 & 18.79$\pm$0.01 & 18.71$\pm$0.01 & 18.81$\pm$0.03 & -- & -- & 15.30$\pm$0.04 & 14.80$\pm$0.06\\
PS1 J024004.78+284027.72 & 4.70 & -27.21 & 20.23$\pm$0.03 & 18.80$\pm$0.01 & 18.78$\pm$0.01 & 18.86$\pm$0.03 & -- & -- & 15.48$\pm$0.04 & 15.09$\pm$0.07\\
PS1 J033236.98+181159.28 & 5.12 & -26.63 & 21.37$\pm$0.05 & 19.92$\pm$0.02 & 19.55$\pm$0.03 & 19.46$\pm$0.04 & 18.39$\pm$0.07 & UHS &16.24$\pm$0.07 & 15.60$\pm$0.13\\
PS1 J034059.31+081115.84 & 4.68 & -27.56 & 19.57$\pm$0.02 & 18.27$\pm$0.01 & 18.53$\pm$0.01 & 18.50$\pm$0.02 & -- & -- & 15.22$\pm$0.04 & 14.77$\pm$0.07\\
PS1 J035643.65+211253.01 & 4.76 & -27.15 & 20.42$\pm$0.02 & 18.93$\pm$0.01 & 18.91$\pm$0.01 & 18.95$\pm$0.03 & -- & -- & 15.70$\pm$0.05 & 15.11$\pm$0.09\\
PS1 J035954.73+054420.19 & 4.82 & -27.73 & 19.96$\pm$0.02 & 18.07$\pm$0.003& 18.25$\pm$0.01 & 18.37$\pm$0.01 & -- & -- & 15.10$\pm$0.04 & 14.49$\pm$0.06\\
PS1 J044432.52--292419.13 & 4.80 & -27.72 & 19.79$\pm$0.02 & 18.34$\pm$0.005& 18.15$\pm$0.02 & 18.41$\pm$0.03 & -- & -- & 15.30$\pm$0.04 & 14.67$\pm$0.05\\
PS1 J045057.37--265541.39 & 4.76 & -27.64 & 20.28$\pm$0.02 & 18.68$\pm$0.004& 18.79$\pm$0.01 & 18.48$\pm$0.02 & -- & -- & 14.85$\pm$0.03 & 14.25$\pm$0.04\\
PS1 J051154.73--075444.30 & 5.12 & -26.53 & 21.70$\pm$0.09 & 20.17$\pm$0.02 & 19.71$\pm$0.03 & 19.57$\pm$0.05 & 18.55$\pm$0.06 & VHS & 16.28$\pm$0.07 & 15.63$\pm$0.11\\
PS1 J061624.38--133806.40 & 5.58 & -27.06 & 22.45$\pm$0.33 & 20.10$\pm$0.03 & 19.33$\pm$0.02 & 19.36$\pm$0.11 & 18.49$\pm$0.06 & VHS & 16.50$\pm$0.08 & 15.58$\pm$0.11\\
PS1 J075745.90--000201.09 & 5.55 & -26.64 & 23.40$\pm$0.24 & 20.81$\pm$0.02 & 19.78$\pm$0.02 & 19.81$\pm$0.06 & 18.92$\pm$0.15 & VHS & 16.20$\pm$0.06 & 15.24$\pm$0.09\\
PS1 J081700.66--163329.29 & 5.06 & -26.88 & 21.65$\pm$0.06 & 19.83$\pm$0.01 & 19.41$\pm$0.02 & 19.19$\pm$0.03 & 18.16$\pm$0.04 & VHS & 15.91$\pm$0.05 & 15.33$\pm$0.09\\
PS1 J091655.68--251145.80 & 4.77 & -29.06 & 18.25$\pm$0.003& 17.01$\pm$0.002& 16.97$\pm$0.002& 17.05$\pm$0.01 & -- & -- & 14.04$\pm$0.03 & 13.40$\pm$0.03\\
PS1 J094146.16--011748.03 & 4.95 & -26.22 & 21.59$\pm$0.06 & 20.17$\pm$0.01 & 19.77$\pm$0.02 & 19.87$\pm$0.07 & 19.08$\pm$0.10 & VHS & 16.39$\pm$0.07 & 15.84$\pm$0.16\\
PS1 J095139.70--274212.45 & 4.80 & -27.70 & 19.92$\pm$0.02 & 18.43$\pm$0.005& 18.33$\pm$0.01 & 18.43$\pm$0.02 & -- & -- & 15.18$\pm$0.04 & 14.72$\pm$0.06\\
PS1 J103436.86--123342.49 & 5.51 & -26.44 & 22.69$\pm$0.15 & 20.97$\pm$0.04 & 19.93$\pm$0.02 & 19.99$\pm$0.07 & 19.16$\pm$0.15 & VHS & 16.87$\pm$0.10 & 15.76$\pm$0.14\\
PS1 J105727.60--175831.70 & 5.34 & -26.41 & 22.27$\pm$0.10 & 20.51$\pm$0.02 & 19.59$\pm$0.02 & 19.88$\pm$0.07 & 18.31$\pm$0.05 & VHS & 16.19$\pm$0.06 & 15.32$\pm$0.09\\
PS1 J110837.58--185408.57 & 4.78 & -27.11 & 20.33$\pm$0.02 & 18.89$\pm$0.01 & 18.86$\pm$0.01 & 19.02$\pm$0.03 & -- & -- & 15.98$\pm$0.05 & 15.58$\pm$0.12\\
PS1 J111054.69--301129.93 & 4.83 & -29.04 & 18.65$\pm$0.01 & 17.42$\pm$0.01 & 17.15$\pm$0.01 & 17.09$\pm$0.02 & -- & -- & 14.26$\pm$0.03 & 13.74$\pm$0.03\\
PS1 J111520.32--193506.24 & 4.67 & -27.62 & 19.76$\pm$0.01 & 18.52$\pm$0.01 & 18.51$\pm$0.01 & 18.46$\pm$0.02 & -- & -- & 15.38$\pm$0.04 & 14.95$\pm$0.07\\
PS1 J112143.65--071839.69 & 5.71 & -26.40 & 24.34$\pm$0.71 & 21.73$\pm$0.06 & 19.87$\pm$0.03 & 20.12$\pm$0.08 & 19.30$\pm$0.23 & VHS & 17.00$\pm$0.12 & 15.77$\pm$0.14\\
PS1 J113530.40--152610.25 & 4.63 & -26.69 & 20.15$\pm$0.01 & 19.21$\pm$0.01 & 19.29$\pm$0.01 & 19.38$\pm$0.04 & -- & -- & 16.51$\pm$0.08 & 16.14$\pm$0.20\\
PS1 J113735.37--104934.54 & 5.50 & -26.68 & 22.84$\pm$0.13 & 20.52$\pm$0.03 & 19.78$\pm$0.02 & 19.75$\pm$0.04 & 18.59$\pm$0.05 & VHS & 16.51$\pm$0.08 & 15.58$\pm$0.12\\
PS1 J121402.71--123548.67 & 4.74 & -27.75 & 19.63$\pm$0.01 & 18.28$\pm$0.005& 18.43$\pm$0.01 & 18.36$\pm$0.02 & -- & -- & 15.63$\pm$0.05 & 15.23$\pm$0.09\\
PS1 J123141.01--184149.86 & 4.78 & -27.33 & 20.76$\pm$0.02 & 18.90$\pm$0.01 & 18.94$\pm$0.01 & 18.79$\pm$0.02 & -- & -- & 15.70$\pm$0.04 & 15.32$\pm$0.09\\
PS1 J125049.27--065758.59 & 4.72 & -27.48 & 20.14$\pm$0.01 & 18.64$\pm$0.01 & 18.59$\pm$0.01 & 18.63$\pm$0.02 & -- & -- & 15.47$\pm$0.04 & 15.08$\pm$0.08\\
PS1 J130031.14--282931.00 & 4.71 & -28.00 & 19.28$\pm$0.01 & 17.89$\pm$0.004& 18.06$\pm$0.01 & 18.09$\pm$0.02 & -- & -- & 14.51$\pm$0.03 & 14.16$\pm$0.04\\
PS1 J131013.11--063951.56 & 5.06 & -26.37 & 21.83$\pm$0.05 & 20.16$\pm$0.02 & 19.91$\pm$0.02 & 19.71$\pm$0.04 & -- & -- & 16.58$\pm$0.08 & 15.98$\pm$0.17\\
PS1 J131326.06--261721.83 & 5.54 & -27.25 & 21.77$\pm$0.08 & 19.91$\pm$0.02 & 19.16$\pm$0.01 & 19.18$\pm$0.03 & 18.18$\pm$0.05 & VHS & 15.90$\pm$0.05 & 14.76$\pm$0.06\\
PS1 J133614.74--183043.29 & 4.77 & -27.39 & 20.26$\pm$0.02 & 18.78$\pm$0.01 & 18.66$\pm$0.01 & 18.72$\pm$0.02 & -- & -- & 15.74$\pm$0.04 & 15.28$\pm$0.09\\
PS1 J140322.58--120905.75 & 4.71 & -26.96 & 20.72$\pm$0.04 & 19.28$\pm$0.01 & 19.18$\pm$0.01 & 19.13$\pm$0.02 & -- & -- & 16.06$\pm$0.06 & 15.58$\pm$0.12\\
PS1 J142721.56--050353.04 & 5.08 & -27.35 & 20.54$\pm$0.02 & 19.29$\pm$0.01 & 18.88$\pm$0.01 & 18.74$\pm$0.02 & 17.90$\pm$0.04 & VHS & 16.02$\pm$0.05 & 15.34$\pm$0.10\\
PS1 J143552.35--123856.17 & 4.81 & -27.04 & 20.98$\pm$0.04 & 19.14$\pm$0.01 & 19.30$\pm$0.02 & 19.08$\pm$0.03 & -- & -- & 15.53$\pm$0.04 & 14.96$\pm$0.06\\
PS1 J151911.32--065042.97 & 4.95 & -26.94 & 20.57$\pm$0.02 & 19.35$\pm$0.01 & 19.36$\pm$0.02 & 19.12$\pm$0.03 & -- & -- & 16.01$\pm$0.06 & 15.31$\pm$0.11\\
PS1 J153241.40--193032.70 & 4.69 & -27.66 & 19.55$\pm$0.01 & 18.33$\pm$0.01 & 18.64$\pm$0.01 & 18.42$\pm$0.01 & -- & -- & 14.87$\pm$0.04 & 14.38$\pm$0.05\\
PS1 J153359.76--181027.21 & 5.03 & -27.39 & 20.95$\pm$0.04 & 19.19$\pm$0.01 & 19.00$\pm$0.01 & 18.67$\pm$0.02 & -- & -- & 15.54$\pm$0.05 & 14.86$\pm$0.07\\
PS1 J172839.85+061509.17\tablenotemark{b}  & 4.63 & -27.91 & 19.38$\pm$0.01 & 18.26$\pm$0.003& 18.16$\pm$0.01 & 18.14$\pm$0.01 & -- & -- & 15.05$\pm$0.04 & 14.57$\pm$0.06\\
PS1 J201939.72--194717.70 & 4.61 & -27.52 & 19.45$\pm$0.01 & 18.47$\pm$0.003& 18.60$\pm$0.01 & 18.54$\pm$0.02 & -- & -- & 14.92$\pm$0.04 & 14.36$\pm$0.06\\
PS1 J202618.71--012256.49 & 4.68 & -27.09 & 19.80$\pm$0.02 & 18.77$\pm$0.01 & 18.84$\pm$0.01 & 18.98$\pm$0.04 & -- & -- & 15.37$\pm$0.04 & 14.88$\pm$0.07\\
PS1 J203310.47+121851.49 & 5.11 & -27.37 & 20.39$\pm$0.02 & 18.97$\pm$0.01 & 18.79$\pm$0.01 & 18.73$\pm$0.02 & -- & -- & 15.62$\pm$0.05 & 15.04$\pm$0.08\\
PS1 J210759.13--032300.53 & 4.64 & -26.92 & 20.26$\pm$0.01 & 19.23$\pm$0.01 & 19.08$\pm$0.01 & 19.15$\pm$0.03 & -- & -- & 16.11$\pm$0.06 & 15.70$\pm$0.13\\
PS1 J214109.55+220206.04 & 4.76 & -28.14 & 19.33$\pm$0.01 & 18.12$\pm$0.003& 18.13$\pm$0.01 & 17.97$\pm$0.01 & -- & -- & 15.18$\pm$0.04 & 14.71$\pm$0.06\\
PS1 J220158.60-202627.36 & 4.73 & -28.10 & 19.65$\pm$0.01 & 18.22$\pm$0.002& 18.13$\pm$0.01 & 18.01$\pm$0.01 & -- & -- & 15.12$\pm$0.04 & 14.63$\pm$0.07\\
PS1 J221413.89--153712.13 & 4.76 & -27.14 & 20.12$\pm$0.01 & 19.13$\pm$0.01 & 18.98$\pm$0.01 & 18.95$\pm$0.03 & -- & -- & 15.83$\pm$0.06 & 15.36$\pm$0.11\\
PS1 J222357.87--252634.24 & 4.80 & -27.37 & 20.25$\pm$0.02 & 18.63$\pm$0.01 & 18.77$\pm$0.01 & 18.77$\pm$0.03 & -- & -- & 15.14$\pm$0.04 & 14.53$\pm$0.06\\
PS1 J225944.26+093624.42 & 4.87 & -27.51 & 20.28$\pm$0.02 & 18.80$\pm$0.01 & 18.76$\pm$0.01 & 18.62$\pm$0.02 & -- & -- & 15.58$\pm$0.04 & 15.08$\pm$0.08\\
PS1 J230126.46+245709.41 & 4.96 & -27.42 & 20.28$\pm$0.02 & 18.96$\pm$0.01 & 18.74$\pm$0.01 & 18.65$\pm$0.02 & -- & -- & 15.43$\pm$0.04 & 14.98$\pm$0.07\\
PS1 J230640.20+264434.91 & 4.64 & -27.79 & 19.33$\pm$0.01 & 18.34$\pm$0.004& 18.30$\pm$0.01 & 18.28$\pm$0.01 & -- & -- & 14.71$\pm$0.03 & 14.35$\pm$0.05\\
 \enddata
\tablenotetext{a}{$J$ band data is only listed here for quasars from $z \sim 5.5$ selection. Other quasars without $J$ band data are from $z \sim 5$ sample.}
\tablenotetext{b}{Quasars J0205+2023 and J1728+0615 were selected before PS1 DR1 (unpublished version) and do not meet PS1 DR1 colors.}
\end{deluxetable*}

\begin{deluxetable*}{l c c l l l l c l}%T5
\tablecaption{Spectroscopic Information of 51 Newly Identified Quasars. \label{tbl-5}}
\tabletypesize{\scriptsize}
\tablewidth{0pt}
\tablehead{
\colhead{Name} & 
\colhead{Redshift} &
\colhead{$M_{1450}$}&
\colhead{Selection}  & \colhead{Instrument} &
\colhead{Exptime(s)}  & \colhead{Grating} &
\colhead{Slit}  & \colhead{ObsDate}
}
\startdata
PS1 J001749.00+271052.10 & 4.76 & -27.18 & z5 & LJ2.4m/YFOSC & 2400.0 & G5 & 1.8 & 2016-12-30\\
PS1 J005243.45+455632.03 & 5.31 & -27.05 & z55 & MMT/Red & 300.0 & G270 & 1.5 & 2017-11-17\\
PS1 J013901.97+412358.00 & 4.64 & -27.19 & z5 & LJ2.4m/YFOSC & 2400.0 & G5 & 1.8 & 2016-12-31\\
PS1 J020523.76+202347.55 & 4.69 & -27.23 & -- & LJ2.4m/YFOSC & 2400.0 & G5 & 1.8 & 2016-12-30\\
PS1 J024004.78+284027.72 & 4.70 & -27.21 & z5 & LJ2.4m/YFOSC & 2400.0 & G5 & 1.8 & 2016-12-31\\
PS1 J033236.98+181159.28 & 5.12 & -26.63 & z55/z5 & MMT/Red & 1200.0 & G270 & 1.25 & 2017-10-22\\
PS1 J034059.31+081115.84 & 4.67 & -27.56 & z5 & LJ2.4m/YFOSC & 1800.0 & G5 & 1.8 & 2016-12-30\\
PS1 J035643.65+211253.01 & 4.76 & -27.15 & z5 & LJ2.4m/YFOSC & 2400.0 & G5 & 1.8 & 2016-12-31\\
PS1 J035954.73+054420.19 & 4.82 & -27.73 & z5 & LJ2.4m/YFOSC & 1800.0 & G5 & 1.8 & 2016-12-31\\
PS1 J044432.52--292419.13 & 4.80 & -27.72 & z5 & SSO2.3m/WiFeS & 1200.0 & R3000 & 1.0 & 2017-03-05\\
PS1 J045057.37--265541.39 & 4.76 & -27.64 & z5 & SSO2.3m/WiFeS & 1800.0 & R3000 & 1.0 & 2017-03-06\\
PS1 J051154.73--075444.30 & 5.13 & -26.53 & z55 & MMT/Red & 1800.0 & G270 & 1.25 & 2017-10-22\\
PS1 J061624.38--133806.40 & 5.58 & -27.06 & z55 & SSO2.3m/WiFeS & 2400.0 & R3000 & 1.0 & 2017-03-05\\
PS1 J075745.90--000201.09 & 5.54 & -26.64 & z55 & SSO2.3m/WiFeS & 2400.0 & R3000 & 1.0 & 2017-03-05\\
PS1 J081700.66--163329.29 & 5.06 & -26.88 & z55 & SSO2.3m/WiFeS & 2400.0 & R3000 & 1.0 & 2017-03-06\\
PS1 J091655.68--251145.80 & 4.77 & -29.06 & z5 & SSO2.3m/WiFeS & 1200.0 & R3000 & 1.0 & 2017-03-06\\
PS1 J094146.16--011748.03 & 4.94 & -26.22 & z55/z5 & SSO2.3m/WiFeS & 2400.0 & R3000 & 1.0 & 2017-03-06\\
PS1 J095139.70--274212.45 & 4.80 & -27.70 & z5 & SSO2.3m/WiFeS & 3600.0 & R3000 & 1.0 & 2017-03-03\\
PS1 J103436.86--123342.49 & 5.51 & -26.44 & z55 & SSO2.3m/WiFeS & 3600.0 & R3000 & 1.0 & 2017-03-06\\
PS1 J105727.60--175831.70 & 5.34 & -26.41 & z55 & SSO2.3m/WiFeS & 2400.0 & R3000 & 1.0 & 2017-03-05\\
PS1 J110837.58--185408.57 & 4.78 & -27.11 & z5 & SSO2.3m/WiFeS & 1800.0 & R3000 & 1.0 & 2015-05-12\\
PS1 J111054.69--301129.93 & 4.83 & -29.04 & z5 & SSO2.3m/WiFeS & 900.0 & R3000 & 1.0 & 2015-05-12\\
PS1 J111520.32--193506.24 & 4.65 & -27.62 & z5 & SSO2.3m/WiFeS & 1800.0 & R3000 & 1.0 & 2015-05-13\\
PS1 J112143.65--071839.69 & 5.71 & -26.40 & z55 & SSO2.3m/WiFeS & 3000.0 & R3000 & 1.0 & 2017-03-05\\
PS1 J113530.40--152610.25 & 4.62 & -26.69 & z5 & SSO2.3m/WiFeS & 1800.0 & R3000 & 1.0 & 2015-05-12\\
PS1 J113735.37--104934.54 & 5.50 & -26.68 & z55 & SSO2.3m/WiFeS & 3000.0 & R3000 & 1.0 & 2017-03-05\\
PS1 J121402.71--123548.67 & 4.74 & -27.75 & z5 & SSO2.3m/WiFeS & 1200.0 & R3000 & 1.0 & 2015-05-12\\
PS1 J123141.01--184149.86 & 4.77 & -27.33 & z5 & SSO2.3m/WiFeS & 1500.0 & R3000 & 1.0 & 2015-05-14\\
PS1 J125049.27--065758.59 & 4.72 & -27.48 & z5 & SSO2.3m/WiFeS & 1200.0 & R3000 & 1.0 & 2015-05-13\\
PS1 J130031.14--282931.00 & 4.71 & -28.00 & z5 & SSO2.3m/WiFeS & 1200.0 & R3000 & 1.0 & 2015-05-13\\
PS1 J131013.11--063951.56 & 5.06 & -26.37 & z5 & SSO2.3m/WiFeS & 2400.0 & R3000 & 1.0 & 2015-05-12\\
PS1 J131326.06--261721.83 & 5.54 & -27.25 & z55 & SSO2.3m/WiFeS & 1800.0 & R3000 & 1.0 & 2015-05-12\\
PS1 J133614.74--183043.29 & 4.75 & -27.39 & z5 & SSO2.3m/WiFeS & 663.0 & R3000 & 1.0 & 2015-05-13\\
PS1 J140322.58--120905.75 & 4.70 & -26.96 & z5 & SSO2.3m/WiFeS & 2400.0 & R3000 & 1.0 & 2015-05-13\\
PS1 J142721.56--050353.04 & 5.08 & -27.35 & z55 & SSO2.3m/WiFeS & 2400.0 & R3000 & 1.0 & 2017-03-06\\
PS1 J143552.35--123856.17 & 4.81 & -27.04 & z5 & SSO2.3m/WiFeS & 2400.0 & R3000 & 1.0 & 2015-05-13\\
PS1 J151911.32--065042.97 & 4.95 & -26.94 & z5 & SSO2.3m/WiFeS & 1800.0 & R3000 & 1.0 & 2015-05-13\\
PS1 J153241.40--193032.70 & 4.68 & -27.66 & z5 & SSO2.3m/WiFeS & 1200.0 & R3000 & 1.0 & 2015-05-12\\
PS1 J153359.76--181027.21 & 5.03 & -27.39 & z5 & SSO2.3m/WiFeS & 1500.0 & R3000 & 1.0 & 2015-05-13\\
PS1 J172839.85+061509.17 & 4.63 & -27.91 & -- & SSO2.3m/WiFeS & 900.0 & R3000 & 1.0 & 2015-07-20\\
PS1 J201939.72--194717.70 & 4.61 & -27.52 & z5 & SSO2.3m/WiFeS & 900.0 & R3000 & 1.0 & 2015-07-20\\
PS1 J202618.71--012256.49 & 4.67 & -27.09 & z5 & Magellan/LDSS3 & 1800.0 & VPH-Red & 1.0\_center & 2018-07-22\\
PS1 J203310.47+121851.49 & 5.11 & -27.37 & z5 & Magellan/LDSS3 & 1200.0 & VPH-Red & 1.25\_center & 2018-07-23\\
PS1 J210759.13--032300.53 & 4.64 & -26.92 & z5 & Magellan/LDSS3 & 1800.0 & VPH-Red & 1.25\_center & 2018-07-22\\
PS1 J214109.55+220206.04 & 4.76 & -28.14 & z5 & Magellan/LDSS3 & 1200.0 & VPH-Red & 1.25\_center & 2018-07-23\\
PS1 J220158.60--202627.36 & 4.73 & -28.10 & z5 & SSO2.3m/WiFeS & 600.0 & R3000 & 1.0 & 2015-07-20\\
PS1 J221413.89--153712.13 & 4.69 & -27.14 & z5 & SSO2.3m/WiFeS & 671.0 & R3000 & 1.0 & 2015-07-23\\
PS1 J222357.87--252634.24 & 4.80 & -27.37 & z5 & Magellan/LDSS3 & 1200.0 & VPH-Red & 1.25\_center & 2018-07-22\\
PS1 J225944.26+093624.42 & 4.86 & -27.51 & z5 & Magellan/LDSS3 & 1200.0 & VPH-Red & 1.25\_center & 2018-07-23\\
PS1 J230126.46+245709.41 & 4.95 & -27.42 & z5 & Magellan/LDSS3 & 1200.0 & VPH-Red & 1.25\_center & 2018-07-23\\
PS1 J230640.20+264434.91 & 4.63 & -27.78 & z5 & Magellan/LDSS3 & 1200.0 & VPH-Red & 1.25\_center & 2018-07-23\\
\enddata
\tablecomments{Quasars J0332+1811 and J0941--0117 were included by both $z \sim 5.5$ and $z \sim 5$ selections.}
\end{deluxetable*}

\begin{figure*}%f8
\centering
\epsscale{1.08}
\plotone{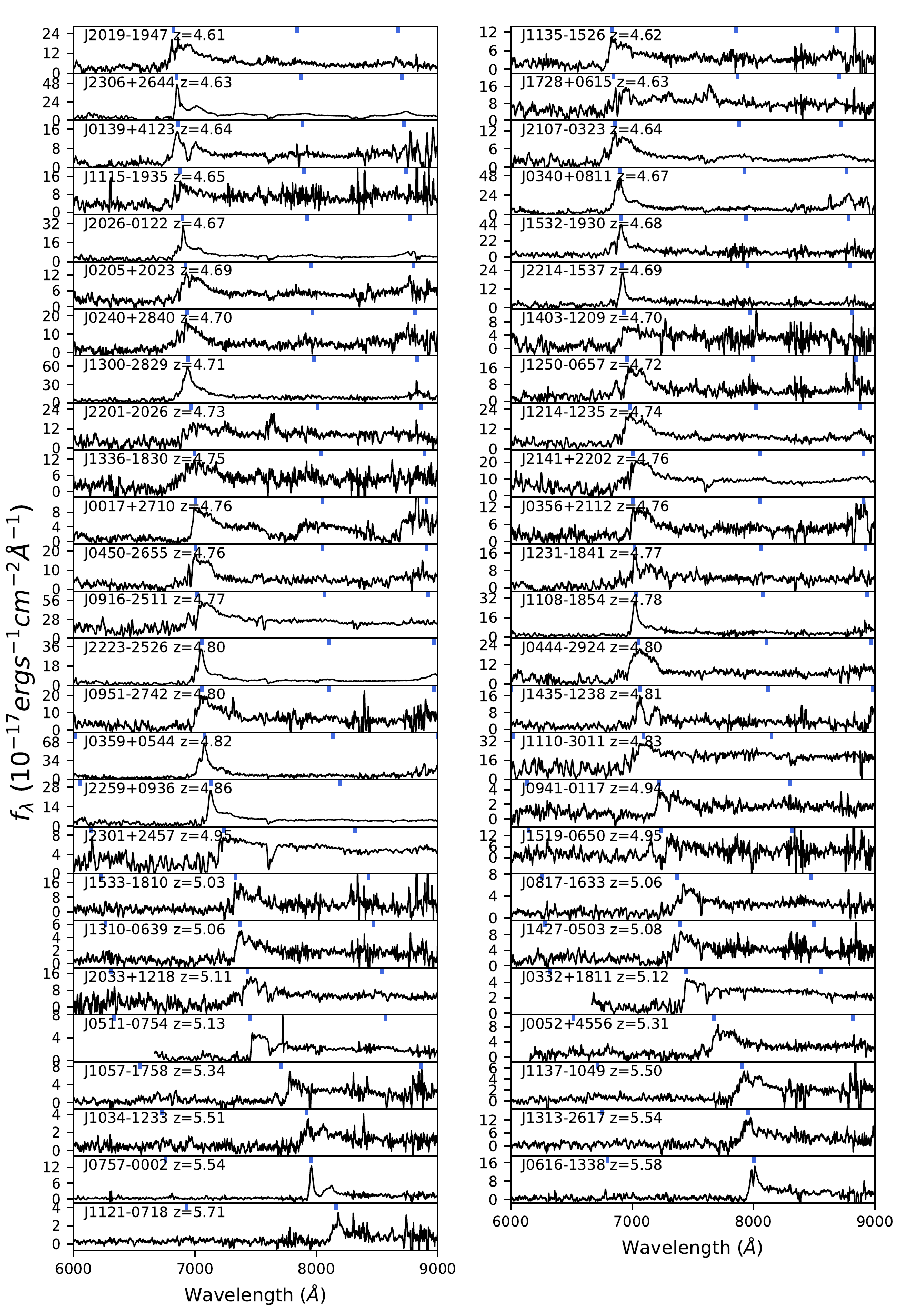}
\caption{The spectra of the 51 newly discovered PS1 selected quasars. The blue vertical lines denote the $\rm Ly \beta$, $\rm Ly \alpha$, Si\,{\sc iv}  and C\,{\sc iv} emission lines. Spectra from ANU 2.3m and Magellan are smoothed with a 10 pixel boxcar. All spectra are corrected for Galactic extinction using the \cite{cardelli89} Milky Way reddening law and E(B $-$ V) derived from the \cite{schlegel98} dust map.}
\end{figure*}

\section{Summary}
Our survey aims at filling the redshift gap of known quasars at $5.3 \le z \le 5.7$, which has been a limitation of the study of IGM evolution, quasar spatial density, and BH evolution over the post-reionization epoch using high redshift quasars. This redshift gap is caused by that quasars at this redshift range have same colors with M dwarfs in broad optical bands. 
To search these quasars, we developed a new selection method and presented our initial discoveries of the $z \sim 5.5$ quasar survey in Paper I. In this paper, we complete the survey and present the final result of our survey. We also apply our method to the PS1 footprint using PS1--J--{\it WISE} dataset and have successfully discovered $\sim 50$ new quasars. Main results of our works are listed as follows.

\begin{itemize}
\item We have completed our $z \sim 5.5$ quasar survey with 15 new quasars reported in this paper. Our survey is based on our new SDSS--PS1--ULAS/UHS/VHS--{\it WISE} selection method and in a $\sim$ 11000 $\rm deg^{2}$ area. This is the first systematic survey of quasars at $z \sim 5.5$. 
\item We have also discovered 31 new quasars at $5.3 \le z \le 5.7$, and other 9 quasars at lower or higher redshift. Our efforts double the number of known quasars in this redshift range. There are 30 new quasars at $5.3 \le z \le 5.7$ in the final SDSS--PS1--ULAS/UHS/VHS--{\it WISE} selected candidate sample. Our survey yields a success rate of 23.1\%. 
\item By combining three previously known quasars and our new discoveries, we construct a quasar sample at $5.2 \le z \le 5.7$ and $M_{1450} < -26.2$. Our sample includes 31 quasars, forming the largest uniformly selected quasar sample at $z \sim 5.5$. We measure the quasar spatial densities at $z = 5.45$ at $M_{1450} < -26.2$ and $M_{1450} < -27$. We compare our measurements with densities at $4 \le z \le 6$ derived from integrating quasar luminosity functions in previous works. Our result ($k= -0.66 \pm 0.05$) shows a good agreement with the rapid decline of quasar spatial density at $z > 5$.
\item We apply our selection pipelines of $z \sim 5$ and 5.5 quasars in PS1-based photometric dataset in the whole PS1 footprint. Using PS1--WISE colors for $z \sim 5$ quasars and PS1--J--{\it WISE} colors for $z \sim 5.5$ quasars, we have discovered 51 new quasars at $4.61 \le z \le 5.71$.
\end{itemize}

Our survey only focuses on bright quasars. So our quasar sample can not reach the faint end of QLF and constrain the spatial density evolution over a wide luminosity range. A larger sample covering wider luminosity range is required for the detailed studies of the QLF and quasar spatial density at $z \sim 5.5$. 
For example, the combination of the PS1, VIKING deep $J$ band photometry and unWISE deep W1 and W2 data \citep{meisner17} will provide dataset for a faint $z \sim 5.5$ quasar survey in $\sim 1500$ deg$^{2}$ area and reach the luminosity $\gtrsim $1.5 magnitude fainter than our sample.  
On the other hand, these bright quasars build a valuable dataset for the study of IGM evolution in the tail of reionization. 
Although the discovery spectra do not have enough signal-to-noise ratio for IGM studies, their high luminosity makes them ideal targets for follow-up deep optical/NIR spectroscopy.
These $z \sim 5.5$ quasars filling the gap of quasar redshift distribution are also powerful probes for the investigation of BH growth from $z \sim 6$ to 5, when the fast SMBH growth phase has been suggested to appear. Due to the low atmosphere transmission between at 18000 -- 19500 $\rm \AA$, quasars at $5.4 < z < 5.8$ need space telescope for the NIR spectroscopy (e.g. JWST) to cover the critical MgII line. 
In addition, in our survey, we only use simple color cuts. Several approaches using probabilistic selections have recently played an important role in quasar selections at multi-redshift ranges \cite[e.g.][]{mortlock12, dipompeo15, schindler18}.
At $z \sim 5.5$, there is no good training sample as most of previously known quasars were located in the right-bottom region at the $riz$ color space due to selection
criteria. Our quasars could provide a new training sample at this redshift for future probabilistic selection.

\acknowledgments
J. Yang, X. Fan, M. Yue, J.-T. Schindler and I. D. McGreer acknowledge the support from the US NSF grant  AST 15-15115.
J. Yang, F. Wang, X.-B. Wu, Q. Yang and L. Jiang thank the supports by the National Key R\&D Program of China (2016YFA0400703) and the National Science Foundation of China (11533001, 11721303). This research uses data obtained through the Telescope Access Program (TAP), which has been funded by the National Astronomical Observatories, Chinese Academy of Sciences, and the Special Fund for Astronomy from the Ministry of Finance in China.
This research uses data obtained through the Telescope Access Program (TAP), which has been funded by the Strategic Priority Research Program "The Emergence of Cosmological Structures" (Grant No. XDB09000000), National Astronomical Observatories, Chinese Academy of Sciences, and the Special Fund for Astronomy from the Ministry of Finance in China. 
We acknowledge the use of the MMT 6.5 m telescope, Palomar Hale 5.2 m telescope, ANU 2.3 m telescope, Magellan 6.5m telescope and Lijiang 2.4 m telescope. Observations obtained with the Hale Telescope at Palomar Observatory were obtained as part of an agreement between the National Astronomical Observatories, Chinese Academy of Sciences, and the California Institute of Technology. This work was partially supported by the Open Project Program of the Key Laboratory of Optical Astronomy, National Astronomical Observatories, Chinese Academy of Sciences. 
We acknowledge the support of the staff of the Lijiang 2.4m telescope. Funding for the telescope has been provided by Chinese Academy of Sciences and the People's Government of Yunnan Province

We acknowledge the use of SDSS photometric data. Funding for SDSS-III has been provided by the Alfred P. Sloan Foundation, the Participating Institutions, the National Science Foundation, and the U.S. Department of Energy Office of Science. The SDSS-III Web site is http://www.sdss3.org/. SDSS-III is managed by the Astrophysical Research Consortium for the Participating Institutions of the SDSS-III Collaboration including the University of Arizona, the Brazilian Participation Group, Brookhaven National Laboratory, University of Cambridge, Carnegie Mellon University, University of Florida, the French Participation Group, the German Participation Group, Harvard University, the Instituto de Astrofisica de Canarias, the Michigan State/Notre Dame/JINA Participation Group, Johns Hopkins University, Lawrence Berkeley National Laboratory, Max Planck Institute for Astrophysics, Max Planck Institute for Extraterrestrial Physics, New Mexico State University, New York University, Ohio State University, Pennsylvania State University, University of Portsmouth, Princeton University, the Spanish Participation Group, University of Tokyo, University of Utah, Vanderbilt University, University of Virginia, University of Washington, and Yale University. This publication makes use of data products from the Wide-field Infrared Survey Explorer, which is a joint project of the University of California, Los Angeles, and the Jet Propulsion Laboratory/California Institute of Technology, and NEOWISE, which is a project of the Jet Propulsion Laboratory/California Institute of Technology. {\it WISE} and NEOWISE are funded by the National Aeronautics and Space Administration. We acknowledge the use of the UKIDSS data, the UHS data, the VISTA data, and PS1 data.

\facility{Sloan (SDSS)}, {WISE}, {UKIRT (WFCam)}, {VISTA}, {Pan-STARR1}, {MMT (Red Channel)}, {Palomar P200/Caltech (DBSP)}, {2.3m/ANU (WiFeS)}, {2.4 m/YAO (YFOSC)}, {Magellan Clay (LDSS3)}.

\end{document}